\documentclass[twocolumn, aps, prb,superscriptaddress]{revtex4}
\usepackage[utf8]{inputenc}
\usepackage{amsmath}
\usepackage{bm}
\usepackage{booktabs}
\usepackage{siunitx}
\usepackage{graphicx}
\usepackage{multirow}
\usepackage{hyperref}
\usepackage{float}
\usepackage[version=4]{mhchem}

\begin{document}

\title{Crystalline electric field excitations in Weyl semimetal \textit{R}AlSi (\textit{R} = Ce, Pr and Nd)}%

\author{Lin Yang}
\affiliation{Beijing National Laboratory for Condensed Matter Physics, Institute of Physics, Chinese Academy of Sciences, Beijing 100190, China}
\affiliation{University of Chinese Academy of Sciences, Beijing 100049, China}

\author{Yili Sun}
\affiliation{Beijing National Laboratory for Condensed Matter Physics, Institute of Physics, Chinese Academy of Sciences, Beijing 100190, China}
\affiliation{University of Chinese Academy of Sciences, Beijing 100049, China}

\author{Xiutong Deng}
\affiliation{Beijing National Laboratory for Condensed Matter Physics, Institute of Physics, Chinese Academy of Sciences, Beijing 100190, China}
\affiliation{University of Chinese Academy of Sciences, Beijing 100049, China}

\author{Weizheng Cao}
\affiliation{School of Physical Science and Technology, ShanghaiTech University, Shanghai201210, China}

\author{Xiaoyan Ma}
\affiliation{Beijing National Laboratory for Condensed Matter Physics, Institute of Physics, Chinese Academy of Sciences, Beijing 100190, China}

\author{Yinguo Xiao}
\affiliation{School of Advanced Materials, Shenzhen Graduate School, Peking University, Shenzhen 518055, China}

\author{Zhentao Wang}
\affiliation{Center for Correlated Matter and School of Physics, Zhejiang University, Hangzhou 310058, China}

\author{Ze Hu}
\affiliation{Spallation Neutron Source Science Center, Dongguan 523803, China}
\affiliation{Institute of High Energy Physics, Chinese Academy of Sciences, Beijing 100049, China}
\affiliation{Guangdong Provincial Key Laboratory of Extreme Conditions, Dongguan 523803, China}

\author{Xiaowen Hao}
\affiliation{Spallation Neutron Source Science Center, Dongguan 523803, China}
\affiliation{Institute of High Energy Physics, Chinese Academy of Sciences, Beijing 100049, China}
\affiliation{Guangdong Provincial Key Laboratory of Extreme Conditions, Dongguan 523803, China}

\author{Yuan Yuan}
\affiliation{Spallation Neutron Source Science Center, Dongguan 523803, China}
\affiliation{Institute of High Energy Physics, Chinese Academy of Sciences, Beijing 100049, China}
\affiliation{Guangdong Provincial Key Laboratory of Extreme Conditions, Dongguan 523803, China}

\author{Zecong Qin}
\affiliation{Spallation Neutron Source Science Center, Dongguan 523803, China}
\affiliation{Institute of High Energy Physics, Chinese Academy of Sciences, Beijing 100049, China}
\affiliation{Guangdong Provincial Key Laboratory of Extreme Conditions, Dongguan 523803, China}

\author{Wei Luo}
\affiliation{Spallation Neutron Source Science Center, Dongguan 523803, China}
\affiliation{Institute of High Energy Physics, Chinese Academy of Sciences, Beijing 100049, China}
\affiliation{Guangdong Provincial Key Laboratory of Extreme Conditions, Dongguan 523803, China}

\author{Qingyong Ren}
\affiliation{Spallation Neutron Source Science Center, Dongguan 523803, China}
\affiliation{Institute of High Energy Physics, Chinese Academy of Sciences, Beijing 100049, China}
\affiliation{Guangdong Provincial Key Laboratory of Extreme Conditions, Dongguan 523803, China}

\author{Xin Tong}
\affiliation{Spallation Neutron Source Science Center, Dongguan 523803, China}
\affiliation{Institute of High Energy Physics, Chinese Academy of Sciences, Beijing 100049, China}
\affiliation{Guangdong Provincial Key Laboratory of Extreme Conditions, Dongguan 523803, China}

\author{Mohamed Aouane}
\affiliation{ISIS Neutron and Muon Source, Rutherford Appleton Laboratory, Chilton, Didcot, OX11 0QX, UK}

\author{Manh Duc Le}
\affiliation{ISIS Neutron and Muon Source, Rutherford Appleton Laboratory, Chilton, Didcot, OX11 0QX, UK}

\author{Youguo Shi}
\email{ygshi@iphy.ac.cn}
\affiliation{Beijing National Laboratory for Condensed Matter Physics, Institute of Physics, Chinese Academy of Sciences, Beijing 100190, China}
\affiliation{University of Chinese Academy of Sciences, Beijing 100049, China}

\author{Yanpeng Qi}
\email{qiyp@shanghaitech.edu.cn}
\affiliation{School of Physical Science and Technology, ShanghaiTech University, Shanghai201210, China}
\affiliation{ShanghaiTech Laboratory for Topological Physics, ShanghaiTech University,Shanghai 201210, China}
\affiliation{Shanghai Key Laboratory of High-resolution Electron Microscopy, ShanghaiTechUniversity, Shanghai 201210, China}

\author{Devashibhai Adroja}
\email{devashibhai.adroja@stfc.ac.uk}
\affiliation{ISIS Neutron and Muon Source, Rutherford Appleton Laboratory, Chilton, Didcot, OX11 0QX, UK}

\author{Huiqian Luo}
\email{hqluo@iphy.ac.cn}
\affiliation{Beijing National Laboratory for Condensed Matter Physics, Institute of Physics, Chinese Academy of Sciences, Beijing 100190, China}


\begin{abstract}

The rare earth intermetallic system \textit{R}Al\textit{X} (\textit{R} = rare earth elements, \textit{X} = Si and Ge) is known to be a promising candidate of magnetic Weyl semimetal. Due to the complex interactions between the rare earth elements and surrounding atoms, as well as hybridization with itinerant electrons, this family likely possesses highly intriguing and novel magnetic structures and thus exhibits dynamic behaviors. We systematically probe polycrystalline samples of \textit{R}AlSi (\textit{R} = La, Ce, Pr and Nd) combining inelastic neutron scattering (INS),  heat capacity and magnetic susceptibility measurements. The INS measurements identify well-resolved crystalline electric field (CEF) excitations at 19.2 and 24.9 meV in CeAlSi, at 5.4 meV in PrAlSi, and at 2.5 and 4.2 meV in NdAlSi. We analyzed the INS data using the corresponding CEF models and determined the CEF parameters and ground state wave functions of \textit{R}AlSi (\textit{R} = Ce, Pr and Nd). Our results suggest strong single-ion anisotropy in their ground states:  $|\pm3/2\rangle$ (94.5\%) in CeAlSi, $|\pm3\rangle$ (99.2\%) in PrAlSi, and $|\pm9/2\rangle$ (76.2\%) in  NdAlSi. Notably, the weaker anisotropy and strong exchange interactions in NdAlSi promote competing magnetic orders and CEF splitting at low  temperature, contrasting with the robust CEF levels in magnetic states of CeAlSi and PrAlSi.

\end{abstract}

\maketitle

\section{Introduction}
Weyl semimetals (WSMs) showing nontrivial topological electronic structures have attracted tremendous interest in condensed matter physics~\cite{armitage2018weyl, weng2016topological, wan2011topological, yan2017topological,weng2015weyl, qi2011topological,soluyanov2015type}. The Weyl nodes hosting as a monopole-like distribution of the Berry curvature, play essential roles in the charge transport related to their quantum states. Typically, Weyl nodes arise from breaking either spatial-inversion symmetry (SIS) or time-reversal symmetry (TRS). While the Weyl fermions have been discovered in  nonmagnetic compounds with broken SIS, magnetic WSMs with broken TRS are expected to exhibit large intrinsic anomalous Hall effects due to strong spin-orbit coupling~\cite{liu2018giant, wang2018large, wang2021intrinsic, bernevig2022progress, li2019intrinsic, liu2021anisotropic,guguchia2020tunable, yan2021weyl, liu2022chiral}. On the other hand, the nontrivial topology may influence the magnetic properties, leading to unique behaviors such as the Kondo effect~\cite{otte2008role, akbari2009theory, fuhrman2015interaction, hillier2012muon, iwasa2023weyl}, quasi-bound states between crystalline electric field (CEF) excitations and phonons~\cite{thalmeier1984theory, adroja2012vibron}, chiral magnetic skyrmions~\cite{bogdanov1989thermodynamically}, and novel magnetic excitations~\cite{kim2012magnetic,itoh2016weyl,jenni2019interplay,cai2020spin}. For instance, the spin waves in magnetic WSMs may be affected by the topology of conduction electrons ~\cite{liu2021spin, itoh2016weyl,jenni2019interplay,cai2020spin,zhang2023chiral,liu2023structural,zou2024experimentally,yang2023topological}.

The rare-earth based compounds \textit{R}Al\textit{X} (\textit{R}=La, Ce, Pr, Nd, Sm, etc., and \textit{X}=Ge, Al) assemble a new family of WSMs with the non-centrosymmetric LaPtSi-type structure, covering diverse magnetism and various Weyl states~\cite{su2021multiple,zhang2020anisotropic,xu2017discovery,ng2021origin,cao2022pressure,chang2018magnetic,yang2021noncollinear, bouaziz2024origin, piva2023topological,sun2021mapping,tzschaschel2024nonlinear,cheng2024tunable, sakhya2023observation,morita2024zone, lou2023signature, lyu2020large, wu2023field, lyu2020nonsaturating,yang2020transition, wang2023quantum,yamada2024nernst,kunze2024optical,bouaziz2024origin,wang2021structure,wang2024spin}. Such non-centrosymmetric magnetic WSMs present rare occasions to investigate the interplay between magnetism and Weyl physics, where both spatial-inversion and time-reversal symmetries are broken.  The \textit{R}AlGe compounds were initially discovered to host various magnetic and WSM states, such as type-II WSM states in the nonmagnetic LaAlGe~\cite{xu2017discovery, ng2021origin}, antimeron pairs in CeAlGe~\cite{suzuki2019singular, puphal2020topological, drucker2023topology,he2023pressure}, and numerous Weyl nodes in the ferromagnetic (FM) PrAlGe~\cite{sanchez2020observation, meng2019large}.  Soon, the \textit{R}AlSi family was also found to have strong interplay between the magnetism and Weyl fermions. Table~\ref{tab:1} summarizes the magnetic properties of \textit{R}AlSi. Specifically, the non-magnetic compound LaAlSi is a potential topological superconductor with $T_c \sim 2.5$~K at 78.5 GPa~\cite{cao2022pressure}. While CeAlSi is a type-I WSM with noncollinear FM order and pressure tunable Weyl nodes~\cite{yang2021noncollinear, bouaziz2024origin, piva2023topological,sun2021mapping,tzschaschel2024nonlinear,cheng2024tunable, sakhya2023observation,morita2024zone}, PrAlSi is a FM type-II WSM with a huge anomalous Hall conductivity of 2000 $\Omega^{-1} \cdot \text{cm} ^{-1}$~\cite{lyu2020large}. For NdAlSi, an incommensurate  ferrimagnetic (FIM) order forms below 7.2 K with a wavevector approximately connected by four pairs of Weyl nodes~\cite{gaudet2021weyl, wang2022ndalsi,lygouras2024magnetic, li2023emergence, zhang2024abnormally}. Recently, possible Weyl-induced spiral order, FM quantum criticality, and altermagnetism are proposed in SmAlSi~\cite{yao2023large,xu2022shubnikov}, EuAlSi~\cite{walicka2024magnetism} and antiferromagnetic (AFM) GdAlSi~\cite{nag2024gdalsi,gong2024magnetic}, respectively. Therefore, the \textit{R}AlSi system is a promising platform for exploring the rich physics in topological electronic structure, static magnetic texture, as well as their dynamic interactions.

In this work, we focus on the study of CEF excitations in \textit{R}AlSi (\textit{R} = Ce, Pr, and Nd) to obtain the electronic ground states of the rare earth ion, from which the typical thermodynamic and magnetic properties of these materials are measured and compared with calculation results. Time-of-flight inelastic neutron scattering (INS) experiments were performed to measure the CEF excitations in polycrystalline samples of \textit{R}AlSi (\textit{R} = Ce, Pr, and Nd) with incident energy $E_{\text{i}}$ up to  180 meV at various temperatures. As CEF excitations are single ion in nature and their $Q$-dependent intensity simply follows the magnetic form factor, the analysis of CEF excitations using  polycrystalline materials is easier than using single crystal study. We observe clear CEF excitations in three magnetic compounds with different energy transfers: $E=$ 19.2 and 24.9 meV in CeAlSi, $E=$ 5.4~meV in PrAlSi, and $E=$ 2.5 and 4.2 meV in NdAlSi. CEF level schemes at different temperatures and the ground states of \textit{R}AlSi were obtained by jointly fitting the INS intensities and the heat capacity data. Our results suggest that the magnetocrystalline anisotropy likely has a significant influence on the magnetic properties of \textit{R}AlSi (\textit{R} = Ce, Pr, and Nd). CeAlSi and PrAlSi exhibit FM ordering driven by strong magnetocrystalline anisotropy, while NdAlSi tends to adopt a complex magnetic structure due to relatively weaker anisotropy. However, at low temperature, the magnetic anisotropy energy in NdAlSi plays a crucial role for the suppression of the incommensurate magnetic order.

 \begin{table}[htbp]
    \centering
    \caption{Magnetic transitions of \textit{R}AlSi}
    \begin{tabular}{*{3}{c}}
        \toprule
          \cline{1-3}
        Materials & Transition temperatures & Magnetic properties \\
        \midrule
         LaAlSi~\cite{cao2022pressure}  & $T_\text{c} \approx 2.5 $ K at 78.5 GPa & Superconducting \\
        CeAlSi~\cite{yang2021noncollinear} & $T_\text{C} \approx 8.2 \text{ K}$   & Noncollinear FM \\
        {PrAlSi~\cite{ lyu2020nonsaturating}} & $T_\text{C} \approx 17.8 \text{ K}$ & {Multiple FM} \\
                                & $T_{M1} \approx 16.5 \text{ K}$ &{Multiple FM}  \\
                                & $T_{M2} \approx 9 \text{ K}$ &{Multiple FM} \\
        {NdAlSi~\cite{gaudet2021weyl}} & $T_\text{inc}\approx 7.2$ K & Incommensurate FIM \\
                                & $T_\text{com} \approx 3.2$ K & Commensurate FIM \\
        {SmAlSi~\cite{yao2023large}} & $T_\text{1}\approx 10.6$ K & {Spiral magnetic order} \\
                                & $T_2 \approx 4.6 $ K & {Spiral magnetic order}\\
        EuAlSi~\cite{walicka2024magnetism} & $T_\text{C} \approx 25.8 \text{ K}$   & Soft FM \\
        GdAlSi~\cite{nag2024gdalsi} & $T_\text{N} \approx 32 \text{ K}$   & AFM or altermagnetic \\
          \cline{1-3}
        \bottomrule
    \end{tabular}
    \label{tab:1}
\end{table}

\section{Experiments}

\subsection{Sample preparation}

The polycrystalline samples of LaAlSi, CeAlSi, PrAlSi, and NdAlSi were prepared by grinding related single crystals, which were finely powdered in an agate mortar for several hours. The single crystals were synthesized by self-flux method~\cite{wu2023field, gaudet2021weyl}. The starting materials were elemental \textit{R} ingot, Al lump and Si powder with composition \textit{R}:Al:Si=1:10:1 (\textit{R} = La, Ce, Pr and Nd), mixed in an alumina crucible. The crucible was placed in a quartz tube,  protected by quartz wool and sealed under vacuum.  The tube was placed in a muffle furnace, heated to \SI{1000}{\celsius} at \SI{3}{\celsius} min$^{-1}$, held for 12 h, and then cooled to \SI{700}{\celsius} at  \SI{2}{\celsius} h$^{-1}$, maintained at \SI{700}{\celsius} for 12 h and then centrifuged to remove the Al flux. We got shining crystals in millimeters size and then ground them to fine powders for INS measurements.

\subsection{Heat capacity and magnetic susceptibility}

The heat capacity of \textit{R}AlSi crystals was measured by a physical properties measurement system Dynacool (Quantum Design, 9 T) using the standard thermal-relaxation method in the temperature range of 2 - 300 K. The magnetic susceptibility was measured by a superconducting quantum interference device magnetometer (Quantum Design, 7 T) with the Helium-3 refrigerator. The heat capacity for each sample was subtracted by the pre-measured background from the sample holder and the addenda. The magnetic heat capacity of CeAlSi, PrAlSi, and NdAlSi was obtained by further subtracting the specific heat of the nonmagnetic LaAlSi, where they have similar phonon contributions.

\subsection{Neutron scattering}

For the INS measurements, we utilized the time-of-flight chopper spectrometer MARI at the ISIS Facility, the Rutherford Appleton Laboratory, UK~\cite{LE2023168646}, and the High Energy Direct Geometry Time-of-flight spectrometer (31113.02.CSNS.HD) at the China Spallation Neutron Source (CSNS) specifically for characterizing NdAlSi in the low-temperature range $T=$ 2.3 K - 15~K. For MARI measurements, experiments were supported by beam time allocation RB2310280 from the Science and Technology Facilities Council
\cite{ding2023crystal}. Polycrystalline samples were placed in an aluminium foil envelope, which were rolled into annular configuration and placed in a 40 mm diameter thin walled aluminium can. The samples were cooled in He-exchange gas using a closed cycle refrigerator  down to the base temperature around 4.5~K. The data were collected with neutrons of multiple incident energies ($E_\text{i}$) simultaneously in a repetition-rate multiplication mode. Since the CEF excitation levels are compound dependent, we performed INS experiments in different setups: CeAlSi (about 13 g), with $E_\text{i}=$ 9, 19, and 70 meV, at $T=$ 5, 15, 60, and 175~K; PrAlSi (about 6 g), with $E_\text{i}=$ 9, 19, and 70 meV, at $T=$ 5, 14, 60, and 150 K; and NdAlSi (about 10 g), with $E_\text{i}=$ 12, 23, and 180 meV, at $T=$ 5, 15, 60, and 175~K. To identify the phonon excitations, we also performed the measurements on the isostructural nonmagnetic compound LaAlSi, with $E_\text{i}=$ 19 and 70 meV, at $T=$ 5 and 175~K. All MARI data were corrected by the detector efficiency determined by the incoherent scattering from a vanadium ring using a white neutron beam. The NdAlSi sample was further measured at CSNS with $E_\text{i}=$ 12 meV at $T=$ 2.3, 4, 6, 8, 10, and  15~K using a similar setup.

\subsection{CEF model}

In the $4f$-electron rare earth systems, the magnetic ions interact with the surrounding crystalline electric field of the ligand atoms based on the single-ion model~\cite{fulde1985magnetic, Jensen1991}. The magnetism of this system has its origin in the angular moment of the $4f$ electrons. Meanwhile, the magnetic ions can also interact with each other through an indirect exchange mediated by Ruderman–Kittel–Kasuya–Yosida (RKKY) interactions.

Above the magnetic ordering temperature, the single-ion model is capable of reproducing the CEF excitations observed by INS, and also the thermodynamic properties such as the temperature dependence of magnetic heat capacity and entropy. The allowed single-ion terms are strongly restricted by the space group symmetry of the crystal structure. Specifically, CeAlSi and NdAlSi crystallize in the non-centrosymmetric space group $I4_1md$ (No.~109) and the magnetic atoms are located at the Wyckoff position 4$a$ (site symmetry 2\textit{mm}) related to $C_{2v}$ point symmetry~\cite{gaudet2021weyl,Pang2022}. PrAlSi has a centrosymmetric structure under the space group $I4_1/amd$ (No.~141), with the Pr$^{3+}$ ion in $D_{2d}$ point symmetry~\cite{lyu2020nonsaturating}. Further details on the crystalline structure of the \textit{R}AlSi system can be found in the references listed in Table I. Considering the point symmetry $C_{2v}$ and $D_{2d}$ of the magnetic ions, the CEF Hamiltonians with the quantization axis along the $c$ axis are given by
\begin{equation}
    \begin{aligned}
    \hat{H}^{{C}_{2v}}_\text{CEF} = & \ {B_2^0} {\hat{O}_2^0} + {B_2^2} {\hat{O}_2^2} +\\
    & \ {B_4^0} {\hat{O}_4^0} + {B_4^2} {\hat{O}_4^2} + {B_4^4} {\hat{O}_4^4} + \\
    & \ {B_6^0} {\hat{O}_6^0} + {B_6^2} {\hat{O}_6^2} + {B_6^4} {\hat{O}_6^4} +
    {B_6^6} {\hat{O}_6^6}
    \end{aligned}
    \label{C2v}
\end{equation}
and
\begin{equation}
    \hat{H}^{{D}_{2d}}_\text{CEF} = {B_2^0} {\hat{O}_2^0} + {B_4^0} {\hat{O}_4^0} +
    {B_4^4} {\hat{O}_4^4} + {B_6^0} {\hat{O}_6^0} + {B_6^4} {\hat{O}_6^4},
    \label{D2d}
\end{equation}
respectively. Here, ${B_m^n}$ and ${\hat{O}_m^n}$ are Stevens parameters and Stevens operators, respectively~\cite{stevens1952matrix, hutchings1964point}. The degeneracy of the total angular momentum $J$ is split into several CEF levels once we consider these single-ion Hamiltonians, and the matrix elements between these states can be easily evaluated.

\begin{table*}[t]
    \caption{Energy levels and associated wave functions determined from the analysis of the INS data of CeAlSi using the $C_{2v}$ CEF model described in the text.}
    \begin{ruledtabular}
    \begin{tabular}{c|cccccc}
    E (meV) &$| -\frac{5}{2}\rangle$ & $| -\frac{3}{2}\rangle$ & $| -\frac{1}{2}\rangle$ & $| \frac{1}{2}\rangle$ & $| \frac{3}{2}\rangle$ & $| \frac{5}{2}\rangle$ \tabularnewline
        \hline
    0.000 & -0.032 & 0.0 & 0.234 & 0.0 & -0.972 & 0.0 \tabularnewline
    0.000 & 0.0 & 0.972 & 0.0 & -0.234 & 0.0 & 0.032 \tabularnewline
    19.160 & -0.823 & 0.0 & 0.545 & 0.0 & 0.158 & 0.0 \tabularnewline
    19.160 & 0.0 & -0.158 & 0.0 & -0.545 & 0.0 & 0.832 \tabularnewline
    25.066 & 0.567 & 0.0 & 0.805 & 0.0 & 0.175 & 0.0 \tabularnewline
    25.066 & 0.0 & -0.175 & 0.0 & -0.805 & 0.0 & -0.567 \tabularnewline
    \end{tabular}\end{ruledtabular}
    \label{CeAlSi:Eigenvectors}
\end{table*}

\begin{figure}[t]
    \centering
    \includegraphics[width=\columnwidth]{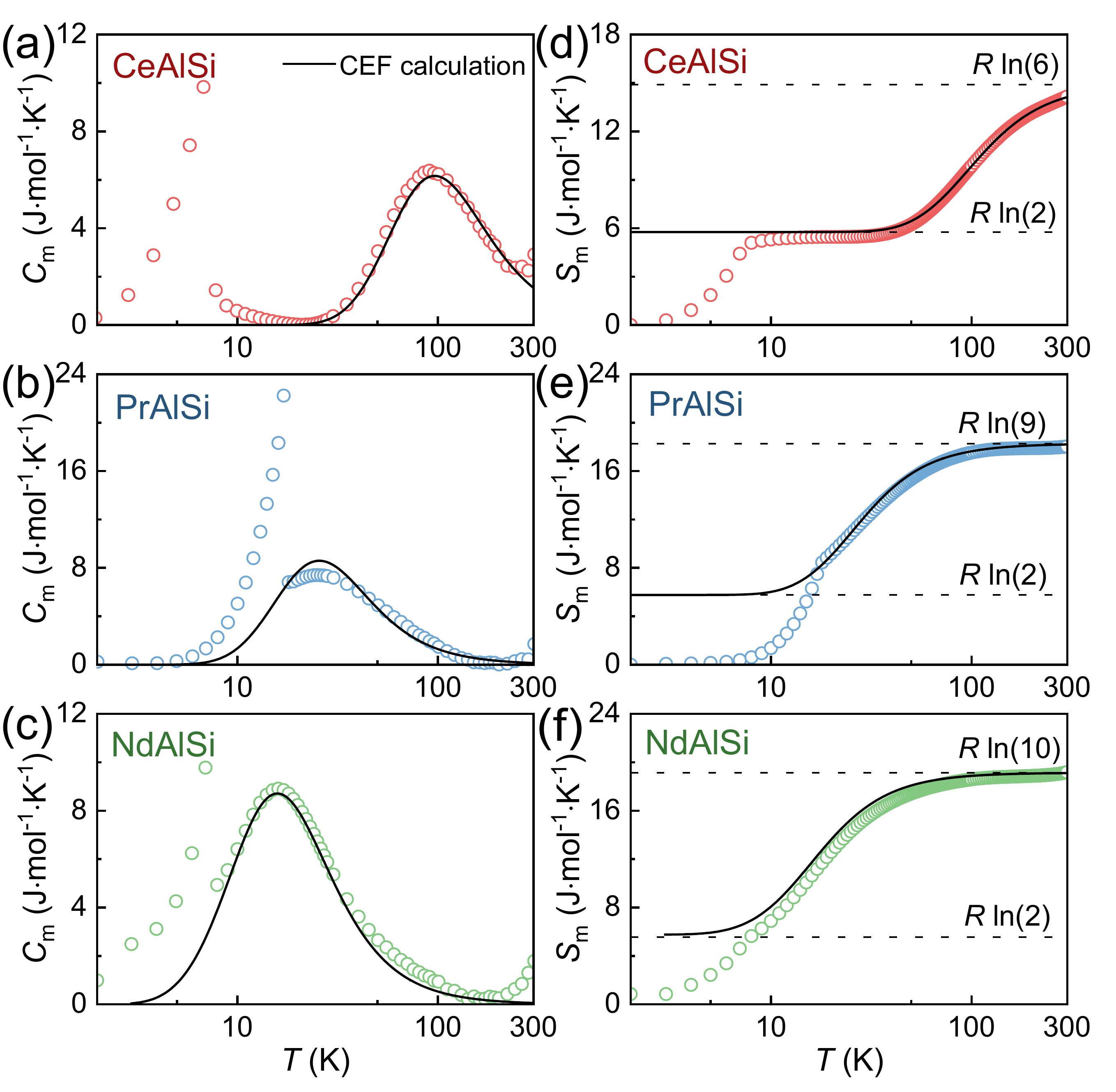}
    \caption{Magnetic heat capacity is plotted as a function of the temperature in (a) CeAlSi, (b) PrAlSi, and (c) NdAlSi, by subtracting the specific heat of LaAlSi. The broad Schottky anomaly is due to CEF excitations, and the sharp peaks are magnetic transitions. Magnetic entropy $S_\text{m}$, obtained by integrating $C_\text{m}/T$ with respect to $T$, is shown as a function of temperature in (d) CeAlSi, (e) PrAlSi, and (f) NdAlSi. All the solid lines are calculated using the related fitted CEF level scheme.}
    \label{fig:Cp}
\end{figure}

\begin{figure}[htbp]
    \centering
    \includegraphics[width=\columnwidth]{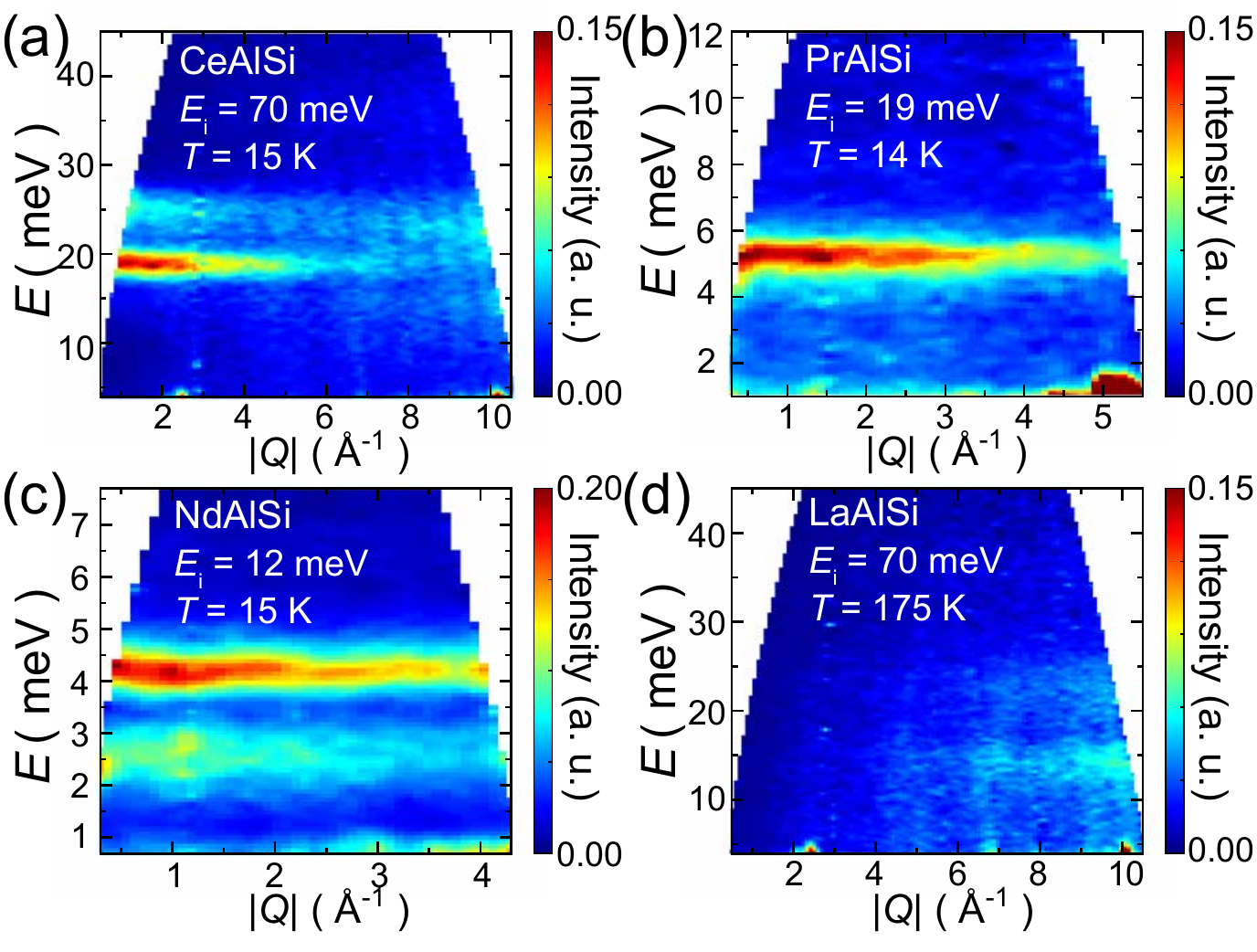}
    \caption{The color-coded map of the inelastic neutron scattering responses in \textit{R}AlSi, plotted as the energy transfer $E$ vs momentum transfer $|\bm{Q}|$ measured of (a) CeAlSi at 15~K with $E_\text{i}=$ 70 meV, (b) PrAlSi at 14~K with $E_\text{i}=$ 19 meV, (c) NdAlSi at 15~K with $E_\text{i}=$ 12 meV,  and (d) LaAlSi at 175~K with $E_\text{i}=$ 70 meV.}
    \label{fig:rawdata}
\end{figure}

At a given wave vector $\bm{Q}$, the neutron cross section associated with the CEF Hamiltonian is~\cite{boothroyd2020principles}
\begin{equation}
    \frac{\text{d}^2\sigma}{\text{d}\Omega \text{d} \omega }
    = A\sum_{m, n} p_n |  \langle \Gamma_m |\hat{J}_{\bot}| \Gamma_n \rangle | ^2
    \delta(\hbar \omega + E_n - E_m),
    \label{eq3}
\end{equation}

\noindent where $A$ is a normalization factor, $p_n$ is the Boltzmann weight, $\hat{J}_{\bot}$ is the component of $\hat{\bm{J}}$ perpendicular to $\bm{Q}$, and $|  \langle \Gamma_m |\hat{J}_{\bot}| \Gamma_n \rangle | ^2$ is computed from the inner product of the matrix element of the magnetic moment with the CEF eigenstates $|\Gamma_n \rangle$.

The CEF excitations and magnetic heat capacity were jointly fitted using a least-squares minimization code, based on the neutron scattering data analysis software Mantid~\cite{arnold2014mantid} and the PyCrystalField CEF calculation package~\cite{scheie2021pycrystalfield}.

\section{Results and Discussions}

The magnetic heat capacity ($C_\text{m}$) of \textit{R}AlSi, as shown in Fig.~\ref{fig:Cp} for (a) CeAlSi, (b) PrAlSi, and (c) NdAlSi was obtained by subtracting the phonon background of the isostructural nonmagnetic LaAlSi. The broad Schottky anomaly observed in $C_\text{m}$ for each sample is attributed to the CEF splitting, and the sharp peak corresponds to the long-range magnetic ordering, consistent with previous reports~\cite{yao2023large}. To better understand the CEF level scheme, the magnetic entropy ($S_\text{m}$) is derived, as shown in Fig.~\ref{fig:Cp}(d)-(f), by integrating $C_\text{m}/T$ with respect to temperature ($T$). The solid lines represent the calculations of both the magnetic specific heat and the entropy based on the CEF states of \textit{R}AlSi. We note that these calculations are valid only above the magnetic transition temperature, as they neglect the exchange interactions that are essential for the formation of long-range magnetic order.

Figure~\ref{fig:rawdata} displays the raw two-dimensional false-color INS spectrum of \textit{R}AlSi, where the INS intensities are plotted as energy transfer $E$ vs the momentum transfer $|\bm{Q}|$. As shown in Fig.~\ref{fig:rawdata}(a)-(c), CEF excitations in CeAlSi, PrAlSi and NdAlSi are only weakly dispersive, which can be seen most clearly at small $|\bm{Q}|$ with decreasing intensity following the magnetic form factors. The increasing intensities of the nonmagnetic compound LaAlSi with $|\bm{Q}|$ [see Fig.~\ref{fig:rawdata}(d)] are neutrons scattered from phonons. This provides a good comparison with magnetic samples. Note that the strong signals starting from $E = $ 0 meV to higher energy transfer at some special small $|\bm{Q}|$ positions (e.g. $|\bm{Q}|\sim $ 1.2 and 2.5 $\text{\AA}^{-1}$) in every spectrum originate from the
tails from the sample Bragg peaks. We notice that the phonon excitations are much weaker than the CEF excitations even at $T=$ 175 K [Fig.~\ref{fig:rawdata}(d)]. As a result, the phonon excitations can be neglected in low-$|\bm{Q}|$ regions, thus it is reasonable to integrate $|\bm{Q}|$ between 0 and 3 $\text{\AA}^{-1}$ to study the CEF excitations in \textit{R}AlSi at low temperatures.

Here we specify the CEF fitting data in different materials in the following subsections.

\subsection{CeAlSi}

\begin{figure}[htbp]
    \centering
    \includegraphics[width=\columnwidth]{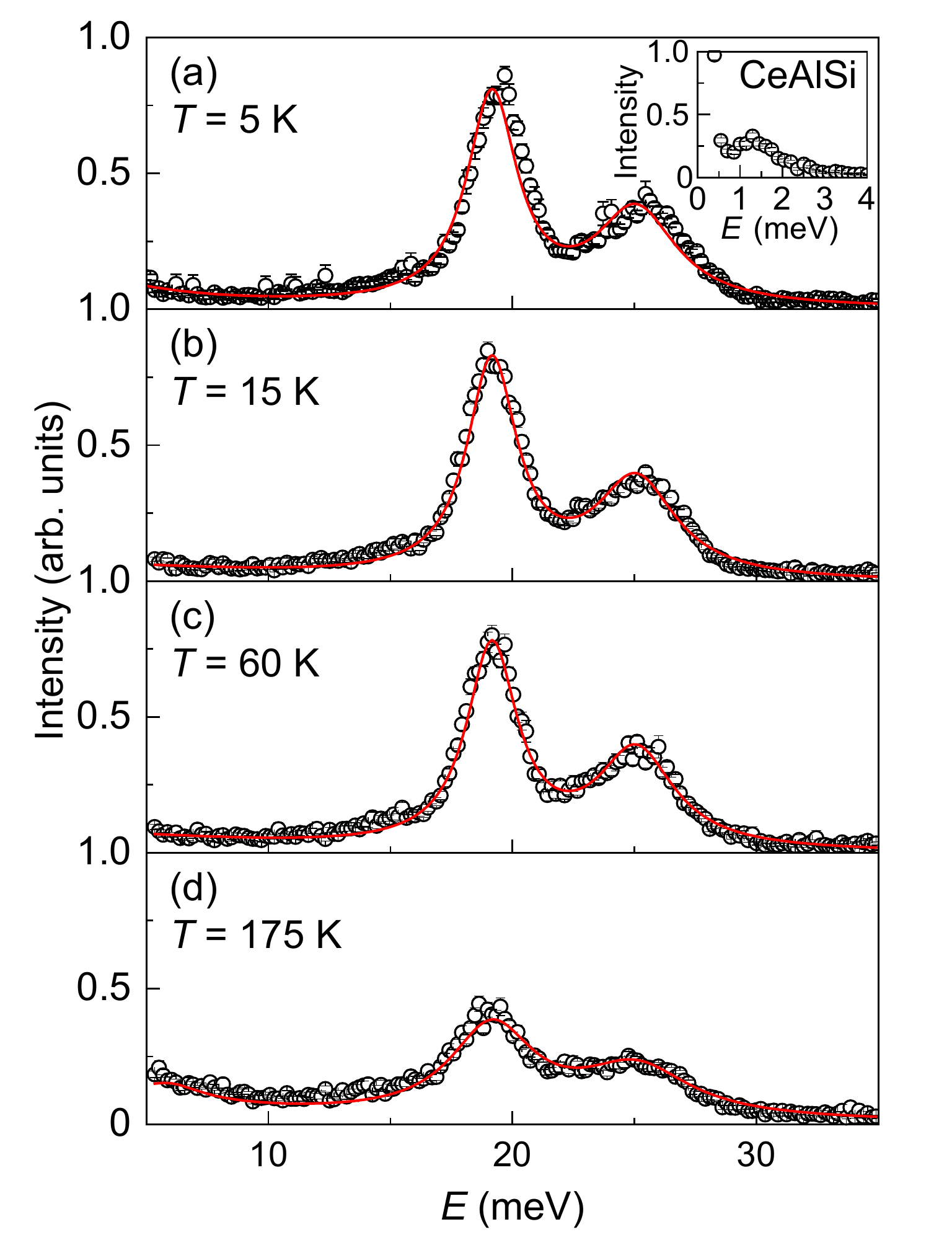}
    \caption{Cuts of the scattering intensity of CeAlSi vs the energy transfer $E$ obtained from integrating across low$-|\bm{Q}|$ ($0-3\text{~\AA}^{-1}$) at different temperatures: $T=$ (a) 5~K, (b) 15~K, (c) 60 K, and (d) 175~K. The solid lines show a fit to the $C_{2v}$ CEF model described in the text. An inset in (a) highlights additional magnetic excitations below 4 meV.}
    \label{fig:CeAlSi}
\end{figure}

\begin{table*}[htbp]
    \caption{Energy levels and associated wave functions determined from the analysis of the INS data of PrAlSi using the $D_{2d}$ CEF model described in the text.}
    \begin{ruledtabular}
    \begin{tabular}{c|ccccccccc}
    E (meV) &$|-4\rangle$ & $|-3\rangle$ & $|-2\rangle$ & $|-1\rangle$ & $|0\rangle$ & $|1\rangle$ & $|2\rangle$ & $|3\rangle$ & $|4\rangle$ \tabularnewline
     \hline
    0.000 & 0.0 & -0.996 & 0.0 & 0.0 & 0.0 & 0.093 & 0.0 & 0.0 & 0.0 \tabularnewline
    0.000 & 0.0 & 0.0 & 0.0 & 0.093 & 0.0 & 0.0 & 0.0 & -0.996 & 0.0 \tabularnewline
    4.743 & 0.0 & 0.0 & -0.707 & 0.0 & 0.0 & 0.0 & 0.707 & 0.0 & 0.0 \tabularnewline
    5.303 & 0.0 & 0.0 & -0.707 & 0.0 & 0.0 & 0.0 & -0.707 & 0.0 & 0.0 \tabularnewline
    5.576 & -0.289 & 0.0 & 0.0 & 0.0 & 0.913 & 0.0 & 0.0 & 0.0 & -0.289 \tabularnewline
    6.638 & 0.0 & 0.093 & 0.0 & 0.0 & 0.0 & 0.996 & 0.0 & 0.0 & 0.0 \tabularnewline
    6.638 & 0.0 & 0.0 & 0.0 & 0.996 & 0.0 & 0.0 & 0.0 & 0.093 & 0.0 \tabularnewline
    9.018 & -0.707 & 0.0 & 0.0 & 0.0 & 0.0 & 0.0 & 0.0 & 0.0 & 0.707 \tabularnewline
    9.707 & -0.645 & 0.0 & 0.0 & 0.0 & -0.409 & 0.0 & 0.0 & 0.0 & -0.645 \tabularnewline
    \end{tabular}\end{ruledtabular}
    \label{PrAlSi:Eigenvectors}
\end{table*}

In CeAlSi, the localized Ce$^{3+}$ ions interact within the non-centrosymmetric lattice that leads to a noncollinear FM spin texture~\cite{yang2021noncollinear} with a sharp FM transition at $T_\text{C}$ = 8.2 K. Since Ce$^{3+}$ is a Kramers ion, the $J$ = 5/2 ground state splits into three CEF doublets in the paramagnetic state under the rhombic CEF potential. The magnetic entropy $S_{\text{m}}$ reaches 5.4 $(\text{J}\cdot \text{mol}^{-1}\cdot \text{K}^{-1})$ at $T_\text{C}$, which is close to the plateaus at $R\ln 2$. This indicates that the magnetic ordering is arising from a Kramers doublet ground state with effective spin-1/2. The magnetic entropy  $S_{\text{m}}$ at 300 K is 14.4 $\text{J}\cdot \text{mol}^{-1}\cdot \text{K}^{-1}$, which is slightly less than $R\ln 6 = 14.9$ $\text{J}\cdot \text{mol}^{-1}\cdot \text{K}^{-1}$ showing the highest CEF levels are close to 300 K (25.9~meV).

The INS measurements detect two well-resolved CEF excitations at around 19.2 and 24.9 meV with incident $E_\text{i}$ = 70 meV, as shown in Fig.~\ref{fig:rawdata}(a) (color map) and Fig.~\ref{fig:CeAlSi} (the circles). As the magnetic entropy displays  plateaus at $R\ln 2$ in a wide temperature range (10 - 50 K), the doublet ground state is well separated from the higher-energy CEF levels in CeAlSi, forming a well-defined effective spin-1/2 model. With incident energy $E_\text{i}$ = 9 meV and $T$ = 5~K ($T < T_C$), an additional excitation around $E$ = 1.5 meV can be observed (see the inset in Fig.3(a)), which is likely from the spin waves of the long-range magnetic order or Zeeman splitting of the ground state doublet under molecular field.

The positions and intensities of the observed CEF excitations in CeAlSi are well described based on the rhombic CEF model Eqs.~\eqref{C2v} and \eqref{eq3} of the $J = 5/2$ multiplet of the Ce$^{3+}$ ions. The $B_6^n$ terms have no effect on CEF splitting for Ce$^{3+}$ ions, hence we set $B_6^n=0$. Based on the joint fit of the high-temperature INS results (at $T=15$, 60, and 175~K, with $E_{\text{i}}=$ 70 meV) and the magnetic specific heat data (10 K$<T<$ 300 K), we obtain $B_2^0=1.928\times10^{-1}$~meV, $B_2^2=1.107\times10^{0}$~meV, $B_4^0=7.281\times10^{-2}$~meV, $B_4^2=-2.335\times10^{-2}$~meV and $B_4^4=0.000$~meV (see Table~~\ref{CeAlSi:Eigenvectors}). The wave eigenfunction of the ground-state doublet ($|m_J\rangle $) is $\psi^{\pm} = \mp0.032|\mp5/2\rangle \pm0.234|\mp1/2\rangle \mp0.972|\pm3/2\rangle$, dominated by the $|\pm3/2\rangle$ states. As shown in Fig.~\ref{fig:CeAlSi}, this set of parameters quantitatively captures the two peaks around 19.2 and 25.1 meV in the INS data. Aside from the redistribution of intensities due to the Bose factor under varying temperatures, there is no significant difference in the INS intensities between the FM state at 5~K and the paramagnetic state at higher temperatures, indicating a relatively weak exchange field in CeAlSi.

\begin{table*}[t]
    \caption{Energy levels and associated wave functions determined from the analysis of the INS data of NdAlSi using the $C_{2v}$ CEF model described in the text.}
    \begin{ruledtabular}
    \begin{tabular}{c|cccccccccc}
    E (meV) &$| -\frac{9}{2}\rangle$ & $| -\frac{7}{2}\rangle$ & $| -\frac{5}{2}\rangle$ & $| -\frac{3}{2}\rangle$ & $| -\frac{1}{2}\rangle$ & $| \frac{1}{2}\rangle$ & $| \frac{3}{2}\rangle$ & $| \frac{5}{2}\rangle$ & $| \frac{7}{2}\rangle$ & $| \frac{9}{2}\rangle$ \tabularnewline
     \hline
    0.000 & 0.0 & -0.010 & 0.0 & -0.190 & 0.0 & 0.217 & 0.0 & -0.394 & 0.0 & -0.873 \tabularnewline
    0.000 & 0.873 & 0.0 & 0.394 & 0.0 & -0.217 & 0.0 & 0.190 & 0.0 & 0.010 & 0.0 \tabularnewline
    2.463 & -0.357 & -0.014 & 0.159 & 0.027 & -0.787 & -0.051 & 0.423 & 0.010 & -0.211 & -0.023 \tabularnewline
    2.463 & -0.023 & 0.211 & 0.010 & -0.423 & -0.051 & 0.787 & 0.027 & -0.159 & -0.014 & 0.357 \tabularnewline
    4.092 & 0.153 & -0.190 & -0.274 & 0.025 & 0.180 & 0.038 & 0.117 & -0.058 & -0.902 & 0.032 \tabularnewline
    4.092 & -0.032 & -0.902 & 0.058 & 0.117 & -0.038 & 0.180 & -0.025 & -0.274 & 0.190 & 0.153 \tabularnewline
    4.445 & 0.125 & -0.020 & -0.145 & -0.118 & -0.492 & -0.070 & -0.826 & -0.021 & -0.140 & 0.018 \tabularnewline
    4.445 & -0.018 & -0.140 & 0.021 & -0.826 & 0.070 & -0.492 & 0.118 & -0.145 & 0.020 & 0.125 \tabularnewline
    6.213 & -0.265 & 0.0 & 0.848 & 0.0 & 0.224 & 0.0 & -0.273 & 0.0 & -0.293 & 0.0 \tabularnewline
    6.213 & 0.0 & 0.293 & 0.0 & 0.273 & 0.0 & -0.224 & 0.0 & -0.848 & 0.0 & 0.265 \tabularnewline
    \end{tabular}\end{ruledtabular}
    \label{NdAlSi:Eigenvectors}
\end{table*}

\subsection{PrAlSi}

\begin{figure}[htbp]
    \centering
    \includegraphics[width=\columnwidth]{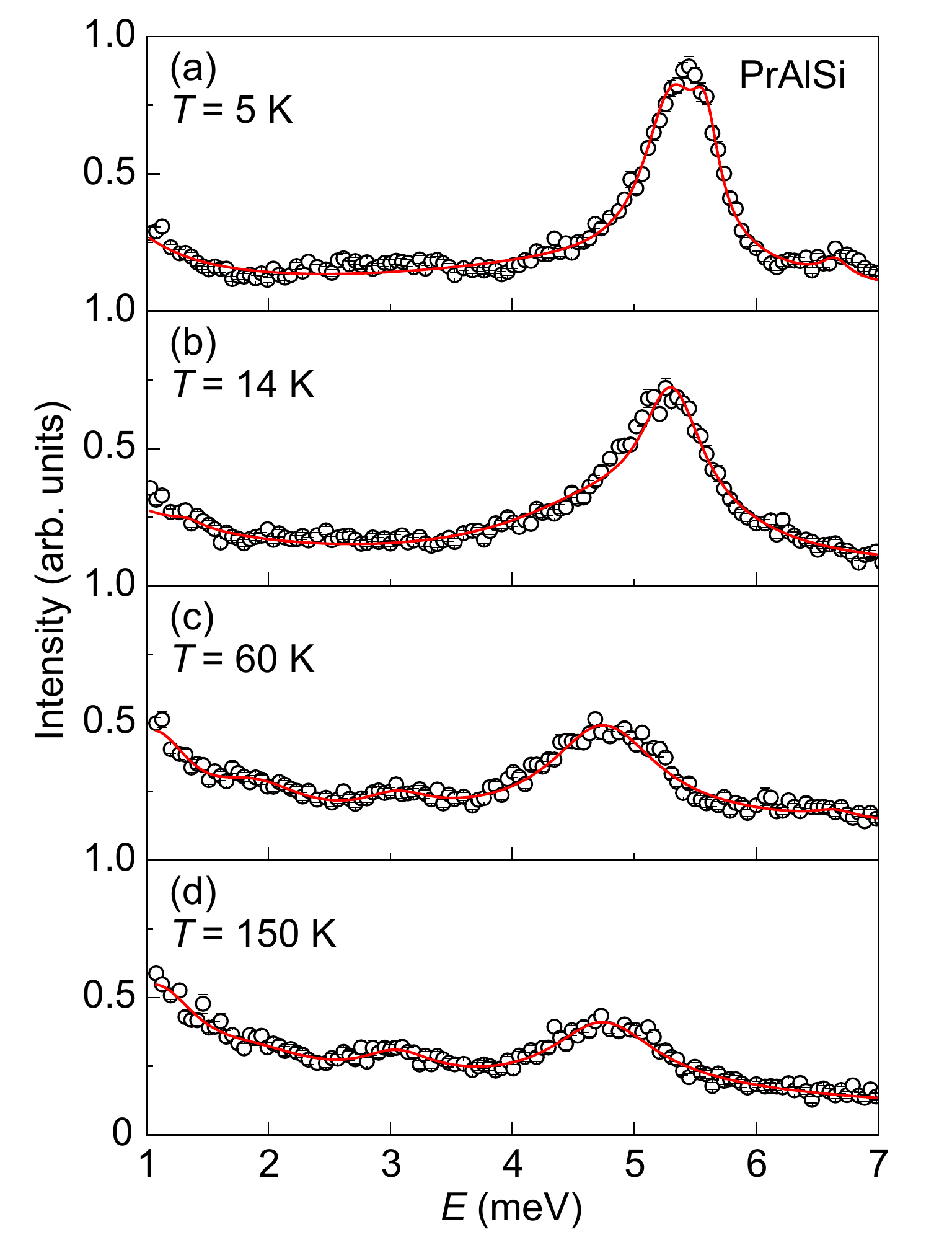}
    \caption{Cuts of the scattering intensity of PrAlSi vs the energy transfer $E$ obtained from integrating across low$-|\bm{Q}|$ ($0-3\text{~\AA}^{-1}$) at different temperatures:  $T=$ (a) 5~K, (b) 14~K, (c) 60 K, and (d) 150 K. The solid lines show a fit to the $D_{2d}$ CEF model described in the text.}
    \label{fig:PrAlSi}
\end{figure}

In PrAlSi, the magnetic susceptibility exhibits a clear FM transition with $T_\text{C} \sim $ 17.8 K, followed by two weak phase transitions at lower temperatures $T_\text{M1} \sim $ 16.5~K and $T_\text{M2} \sim $ 9 K, possibly related to topological magnetic textures or spin glass states~\cite{lyu2020nonsaturating}. Given that Pr$^{3+}$ is a non-Kramers ion with a $J = 4$ ground state multiplet, determining the correct CEF eigenvalues and eigenvectors is challenging if we only used the INS data. The heat capacity measurements of PrAlSi provide critical insight into the CEF distributions. As shown in Fig.~\ref{fig:Cp}(e), the magnetic entropy crosses the $R\ln 2$ line around 13 K (below $T_\text{C}$) and shows no signature of plateau-like behavior. This suggests that the magnetic order in PrAlSi originates not only from the ground state but also from admixture of higher-energy level at finite temperature, potentially related to the two weak phase transitions, $T_\text{M1}$ and $T_\text{M2}$. The magnetic entropy $S_{\text{m}}$ with the value of 17.7 $\text{J}\cdot \text{mol}^{-1}\cdot \text{K}^{-1}$ saturates to the $R\ln 9$ plateau around 100 K, indicating that all CEF levels of $J=4$ lie below 10 meV.

INS measurements were performed on PrAlSi at $T=$ 5~K ($T < T_\text{M2}$), $T=$ 14~K ($T_\text{M2} < T < T_\text{M1}$), and $T=$ 60 K and 150 K (above $T_\text{C}$) with $E_\text{i}=$ 12 meV. Figure~\ref{fig:PrAlSi} shows corresponding one-dimensional cuts of the intensities integrated over $|\bm{Q}|$ ($0-3~\text{\AA}^{-1}$). As shown in Fig.~\ref{fig:rawdata}(b), only one well-defined CEF excitation at 5.4~meV was detected, which is much smaller compared to \textit{LS} coupling energy. This excitation is sharp at $T=$ 5~K ($T < T_\text{M2}$) and less clear at $T=$ 14~K ($T_\text{M2} < T < T_\text{M1}$), where the weak molecular field is still effective. The analysis with Eq.~(\ref{D2d}) confirms the non-Kramers nature of the Pr$^{3+}$ CEF states, where the first excited state is a singlet at 4.7 meV. The corresponding CEF parameters are $B_2^0=2.755\times 10^{-2}$\text{~meV}, $B_4^0=3.087\times 10^{-3}$\text{~meV}, $B_4^4=4.512\times 10^{-3}$\text{~meV}, $B_6^0=7.176\times 10^{-5}$\text{~meV}, and $B_6^4=2.113\times 10^{-4}$\text{~meV}, based on the INS results (at $T=$ 14~K, 60 K and 150~K, with $E_{\text{i}}=$ 19 meV) and the magnetic specific heat data (20 K $<T<$ 300 K). Table.~\ref{PrAlSi:Eigenvectors} displays the calculated eigenvalues and eigenvectors. Similarly to the case of CeAlSi, the ground state wave functions are dominated by the $\psi^{\pm} = 0.093|\pm1\rangle -0.996|\mp3\rangle$ states.

\begin{figure}[htbp]
    \centering
    \includegraphics[width=\columnwidth]{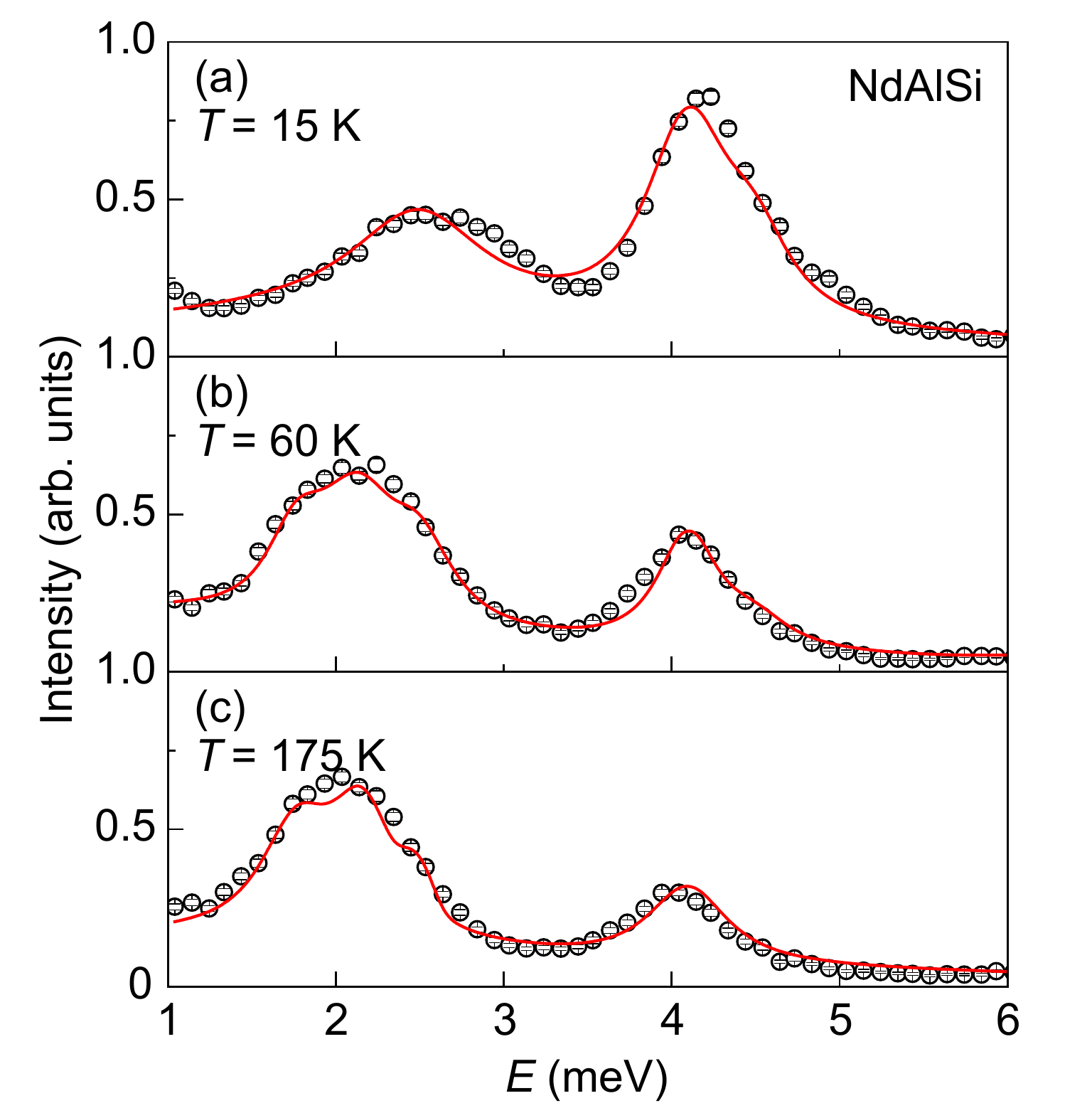}
    \caption{The plots of integrated $|\bm{Q}|$  ($0-3~\text{\AA}^{-1}$) intensity vs the energy transfer $E$ for NdAlSi at different temperatures:  $T=$ (a) 15~K, (b) 60 K and (c) 175~K. The red solid line shows a fit based on the rhombic $C_{2v}$ CEF model.}
    \label{fig:NdAlSi}
\end{figure}

\begin{figure}[htbp]
    \centering
    \includegraphics[width=\columnwidth]{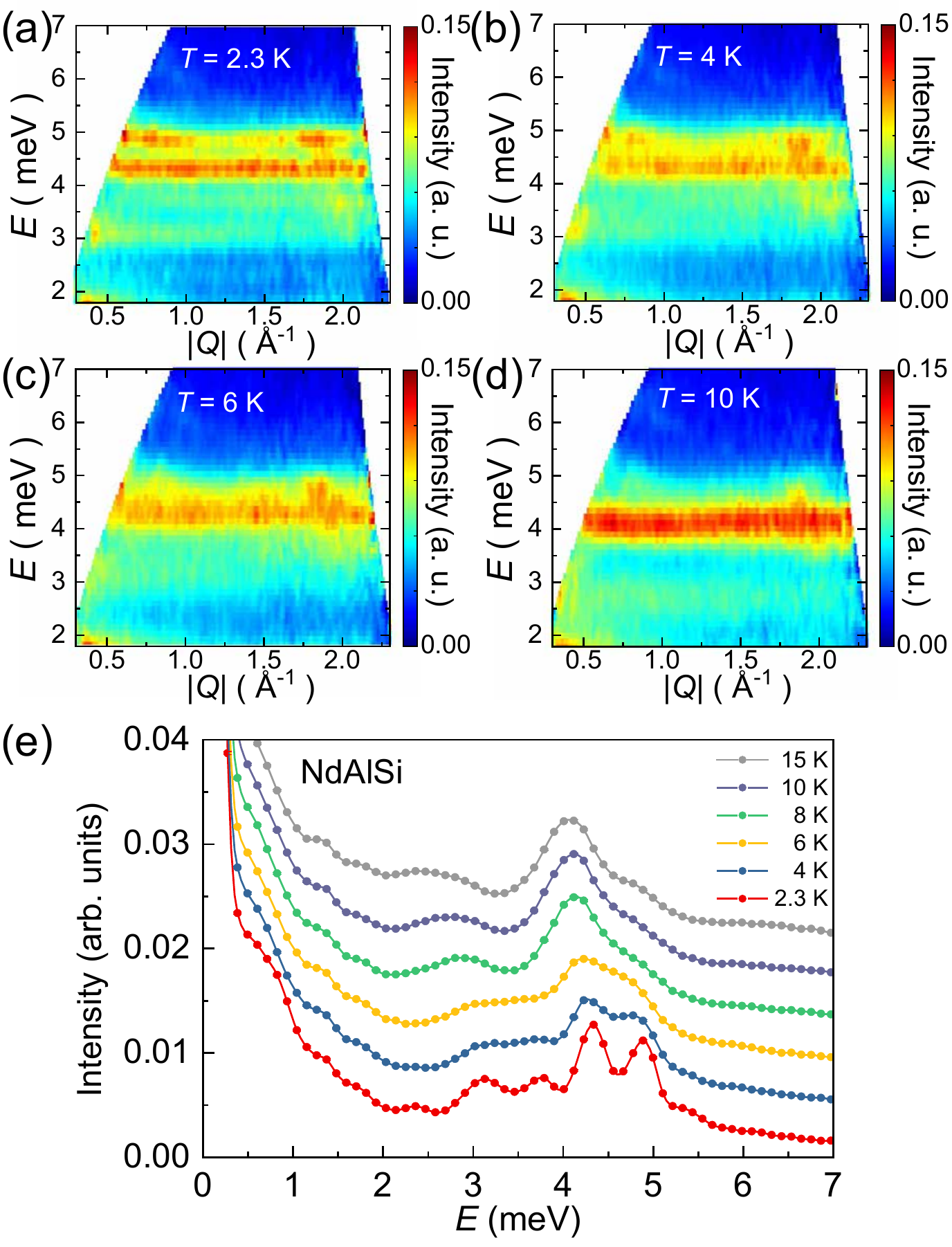}
    \caption{The color-coded map of the INS responses in NdAlSi measured at CSNS, plotted as the energy transfer $E$ vs momentum transfer $|\bm{Q}|$ measured with $E_\text{i}=$ 12 meV at $T=$ (a) 2.3~K, (b) 4~K, (c) 6~K, and (d) 10~K. (e) The plots of integrated $|\bm{Q}|$  ($0-2~\text{\AA}^{-1}$) intensity vs energy transfer $E$ for NdAlSi at different temperatures. Please note that each curve is vertically shifted by 0.005 for clarity.}
    \label{fig:NdAlSi_CSNS}
\end{figure}

\subsection{NdAlSi}

\begin{figure*}[htbp]
    \centering
    \includegraphics[width=2.05\columnwidth]{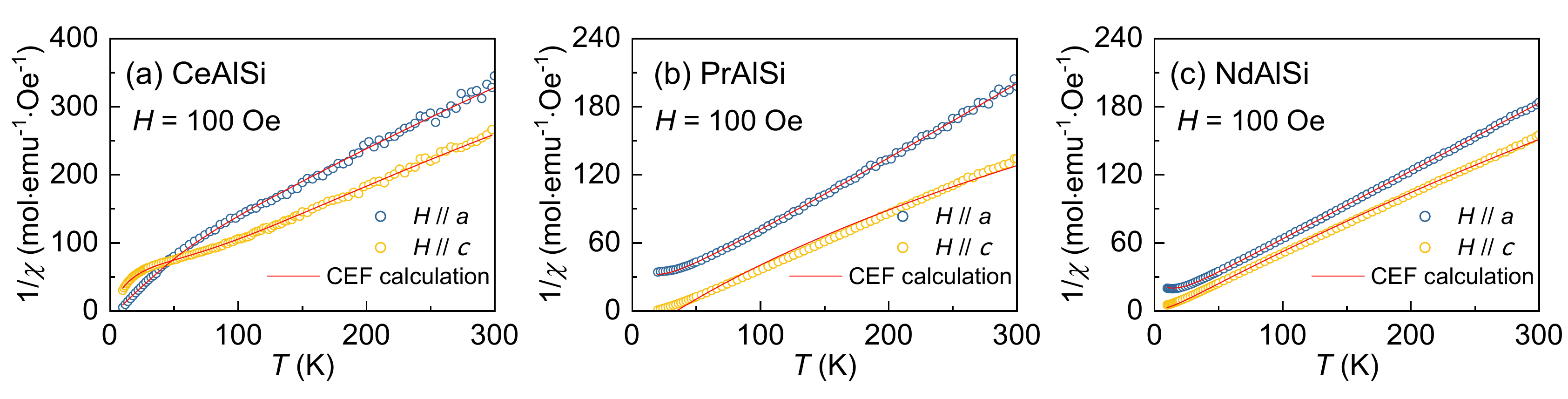}
    \caption{Single crystal magnetic susceptibility of (a) CeAlSi, (b) PrAlSi, and (c) NdAlSi. The experimental data are shown by circles and the solid lines show the fit based on the corresponding CEF model described in the text.}
    \label{fig:sus}
\end{figure*}

Among the \textit{R}AlSi compounds, NdAlSi is special in its complex one-dimensional chiral magnetic structure, where the incommensurate magnetic wavevector $\bm{Q}_\text{inc}$ is found to connect approximately four pairs of Weyl nodes~\cite{gaudet2021weyl}.  The leading instability of NdAlSi is the long-wavelength incommensurate order (below $T_\text{inc} \sim $  7.2 K), followed by a lower temperature (at $T_\text{com} \sim $ 3.3 K) transition to commensurate ferrimagnetism. The $J = 9/2$ multiplet of Nd$^{3+}$ splits into five CEF doublets under the rhombic CEF potential. The magnetic entropy reaches 5.70 $\text{J}\cdot \text{mol}^{-1}\cdot \text{K}^{-1}$ at $T_\text{inc}$ [Fig.~\ref{fig:Cp}(f)], very close to $R\ln 2$ = 5.76 $\text{J}\cdot \text{mol}^{-1}\cdot \text{K}^{-1}$. This indicates that the magnetic ordering arises mostly from the doublet ground state. The value of the magnetic entropy at 300 K is 19.07 $\text{J}\cdot \text{mol}^{-1}\cdot \text{K}^{-1}$, which is slightly less than $R\ln 10$ = 19.14 $\text{J}\cdot \text{mol}^{-1}\cdot \text{K}^{-1}$ expected from the five CEF doublets. This result suggests that the overall CEF level from $J=9/2$ is less than 300 K (25.9 meV).

As shown in Fig.~\ref{fig:NdAlSi}, only two distinct peaks are detected at 2.5 meV and 4.2 meV with incident $E_\text{i} =$ 12 meV at different temperatures. No pronounced magnetic excitations above 5 meV can be detected, even with the highest incident energy $E_\text{i} =$  180 meV. Using  Stevens-operator equivalent method, combining the INS results (at $T=$ 15~K, 60 K and 175~K, with $E_{\text{i}}=$ 12 meV) and the magnetic specific heat data (10 K $<T<$ 300 K), the resulting fit curves (red lines) and  eigenvalues and eigenvectors are shown in Fig.~\ref{fig:NdAlSi} and Table~\ref{NdAlSi:Eigenvectors}, respectively. The CEF parameters of NdAlSi are $B_2^0=-3.909\times 10^{-2}$\text{~meV}, $B_2^2=3.129\times 10^{-2}$\text{~meV}, $B_4^0=-7.703 \times 10^{-4}$\text{~meV}, $B_4^2=-3.477\times 10^{-3}$\text{~meV}, $B_4^4=-2.080\times 10^{-3}$\text{~meV}, $B_6^0={3.402\times 10^{-6}} \text{~meV}$, $B_6^2={-1.331\times 10^{-4}} \text{~meV}$, $B_6^4={6.643 \times 10^{-5}} \text{~meV}$, and $B_6^6={-9.551 \times 10^{-5}} \text{~meV}$. The wave function of the lowest doublet is $\mp0.010|\mp7/2\rangle \mp0.190|\mp3/2\rangle\pm0.217|\pm1/2\rangle\mp0.394|\pm5/2\rangle\mp0.873|\pm9/2\rangle$. The combination of $|m_J\rangle $ indicates weaker magnetic anisotropy compared to CeAlSi and PrAlSi, aligning with the experimental observation of the chiral spin texture in NdAlSi. Due to the limited resolution of the spectrometer, the two excited levels at around 4.2 meV merge into a single peak and the intensity at 6.2 meV is indistinguishable.

Unlike CeAlSi and PrAlSi, which show thermal population changes without observable peak shifts or broadening where the magnetic order develops, NdAlSi demonstrates clear CEF spectrum evolution below 15~K. Magnetic coupling between \ce{Nd^{3+}} ions may result from two contributions: short-range interactions of probably a super-exchange origin and long-range RKKY interactions~\cite{lygouras2024magnetic}.  Fig.~\ref{fig:NdAlSi_CSNS}(a)-(d) show the temperature dependent CEF excitations as 2D slices with $E$ vs  $|\bm{Q}|$ in NdAlSi further measured at CSNS. For clarity, the temperature dependence of the CEF excitations is plotted as the integrated intensity over $|\bm{Q}|$ (0–2~$\text{\AA}^{-1}$) in Fig.~\ref{fig:NdAlSi_CSNS}(e), where each curve is vertically offset by 0.005. In the incommensurate magnetic order state ($T=$ 6 and 4 K), the CEF peaks exhibit broadening along with shifts in their energy levels. Moreover, the magnetic excitation near 3 meV becomes notably diffuse, suggesting that competing magnetic coupling within NdAlSi leads to a metastable magnetic phase, which may be connected to Weyl-mediated RKKY interactions\cite{gaudet2021weyl, zhou2025weyl}. In contrast, in the commensurate magnetic order state ($T=$ 2.3 K), the CEF levels show pronounced splitting. These observations indicate that the commensurate magnetic order in NdAlSi is governed by strongly localized magnetic interactions with the molecular field estimated to be approximately 1.5 T within the plane and 3.7 T out of the plane in NdAlSi according to McPhase\cite{rotter2004using}. Although the static in-plane component of the ordered moment is small\cite{gaudet2021weyl}, our analysis reveals that dynamic in-plane interactions play a crucial role in the splitting of the excited CEF levels, while out-of-plane magnetic interactions dominate the ground state splitting and shifts of excited CEF levels. Therefore, CeAlSi, PrAlSi and NdAlSi have very different ground states due to their distinct spin anisotropy at low temperature\cite{bouaziz2024origin, luo2025correlated}. Such contrast highlights how CEF-driven anisotropy governs not only the local moment alignment but also the collective magnetic behavior, which probably in turn influences the coupling between magnetism and Weyl fermions. There might be weaker anisotropy in heavy rare-earth compounds, causing rich magnetic states in SmAlSi~\cite{yao2023large}, EuAlSi~\cite{walicka2024magnetism}, GdAlSi~\cite{nag2024gdalsi}, etc. However, considering high neutron absorption cross sections of Sm, Eu, and Gd, inelastic neutron scattering study of these compounds is very difficult.

To further test the obtained CEF parameters, we calculated the susceptibility and compared with the experimental results, as shown in Fig.~\ref{fig:sus} for (a) CeAlSi, (b) PrAlSi and (c) NdAlSi. Both the results along the crystallographic $a$ and $c$ directions are in good agreement with the experimental data above the magnetic transition temperature, where $1/\chi$ is nearly linear from 100~K to 300~K. As the temperature approaches the magnetic transition temperatures, some discrepancies emerge likely due to the effect of short-range magnetic coupling.

\section{conclusions}

We have systematically probed the CEF energy levels in the magnetic WSM candidates \textit{R}AlSi (\textit{R} = Ce, Pr and Nd) using heat capacity and INS measurements. The INS measurements confirm the presence of well-resolved CEF excitations in CeAlSi, PrAlSi and NdAlSi.  The CEF fitting results reveal different doublet ground states in CeAlSi, PrAlSi, and NdAlSi arising from their ligand environments which are consistent with the heat capacity measurements. All three compounds exhibit strong magnetic anisotropy, though NdAlSi shows reduced anisotropy compared to CeAlSi and PrAlSi. The ground states of CeAlSi and PrAlSi are predominantly composed of the states $|\pm3/2\rangle$ (94.5\%) and $|\pm3\rangle$ (99.2\%), respectively, consistent with their FM structures. In contrast, NdAlSi displays more complex ground state wave functions: while $|\pm9/2\rangle$ states play a dominant role (76.2\%), other components also contribute, which is consistent with the observed FIM structure. This difference  in NdAlSi likely originates from Dzyaloshinskii-Moriya interactions, where the in-plane spin component plays a significant role. The intermediate anisotropy scale in NdAlSi, evidenced by its wave functions mixing, may enable the coexistence of competing interactions that drive its complex magnetic phase diagram.

Our findings highlight the magnetic anisotropy under CEF effects in \textit{R}AlSi, which constrains the magnetic moments in aligned moment configurations. Compared to CeAlSi and PrAlSi, the slightly weaker anisotropy in NdAlSi may tolerate the incommensurate AFM order while still playing a crucial role in suppressing it at low temperature. Further INS measurements on single crystals are highly desired to resolve the spin waves and determine the spin-spin exchange couplings in \textit{R}AlSi~\cite{lygouras2024magnetic}.

\begin{acknowledgments}
This work is supported by the National Key Research and Development Program of China (Grants No.~2022YFA1403800, No.~2022YFA1403400, No.~2023YFA1406100, No.~2024YFA1408303), the National Natural Science Foundation of China (Grants No.~12274444, No.~11974392, No.~12061130200, No.~12474024, No.~12425512 and No.~12374124), the Chinese Academy of Sciences (Grants No.~XDB33000000, No.~GJTD-2020-01). L.Y., D.T.A. and H.L. would like to thank the Royal Society of London for the funding support of the Newton Advanced Fellowship (Grant No.~NAF/R1/201248).

\end{acknowledgments}


\bibliography{export}

\begin{thebibliography}{86}
\expandafter\ifx\csname natexlab\endcsname\relax\def\natexlab#1{#1}\fi
\expandafter\ifx\csname bibnamefont\endcsname\relax
  \def\bibnamefont#1{#1}\fi
\expandafter\ifx\csname bibfnamefont\endcsname\relax
  \def\bibfnamefont#1{#1}\fi
\expandafter\ifx\csname citenamefont\endcsname\relax
  \def\citenamefont#1{#1}\fi
\expandafter\ifx\csname url\endcsname\relax
  \def\url#1{\texttt{#1}}\fi
\expandafter\ifx\csname urlprefix\endcsname\relax\def\urlprefix{URL }\fi
\providecommand{\bibinfo}[2]{#2}
\providecommand{\eprint}[2][]{\url{#2}}

\bibitem[{\citenamefont{Armitage et~al.}(2018)\citenamefont{Armitage, Mele, and
  Vishwanath}}]{armitage2018weyl}
\bibinfo{author}{\bibfnamefont{N.~P.} \bibnamefont{Armitage}},
  \bibinfo{author}{\bibfnamefont{E.~J.} \bibnamefont{Mele}}, \bibnamefont{and}
  \bibinfo{author}{\bibfnamefont{A.}~\bibnamefont{Vishwanath}},
  \bibinfo{journal}{Rev. Mod. Phys.} \textbf{\bibinfo{volume}{90}},
  \bibinfo{pages}{015001} (\bibinfo{year}{2018}).

\bibitem[{\citenamefont{Weng et~al.}(2016)\citenamefont{Weng, Dai, and
  Fang}}]{weng2016topological}
\bibinfo{author}{\bibfnamefont{H.}~\bibnamefont{Weng}},
  \bibinfo{author}{\bibfnamefont{X.}~\bibnamefont{Dai}}, \bibnamefont{and}
  \bibinfo{author}{\bibfnamefont{Z.}~\bibnamefont{Fang}}, \bibinfo{journal}{J.
  Phys.: Condens. Matter} \textbf{\bibinfo{volume}{28}},
  \bibinfo{pages}{303001} (\bibinfo{year}{2016}), ISSN
  \bibinfo{issn}{0953-8984}.

\bibitem[{\citenamefont{Wan et~al.}(2011)\citenamefont{Wan, Turner, Vishwanath,
  and Savrasov}}]{wan2011topological}
\bibinfo{author}{\bibfnamefont{X.}~\bibnamefont{Wan}},
  \bibinfo{author}{\bibfnamefont{A.~M.} \bibnamefont{Turner}},
  \bibinfo{author}{\bibfnamefont{A.}~\bibnamefont{Vishwanath}},
  \bibnamefont{and} \bibinfo{author}{\bibfnamefont{S.~Y.}
  \bibnamefont{Savrasov}}, \bibinfo{journal}{Phys. Rev. B}
  \textbf{\bibinfo{volume}{83}}, \bibinfo{pages}{205101}
  (\bibinfo{year}{2011}).

\bibitem[{\citenamefont{Yan and Felser}(2017)}]{yan2017topological}
\bibinfo{author}{\bibfnamefont{B.}~\bibnamefont{Yan}} \bibnamefont{and}
  \bibinfo{author}{\bibfnamefont{C.}~\bibnamefont{Felser}},
  \bibinfo{journal}{Annu. Rev. Condens. Matter Phys.}
  \textbf{\bibinfo{volume}{8}}, \bibinfo{pages}{337} (\bibinfo{year}{2017}),
  ISSN \bibinfo{issn}{1947-5454, 1947-5462}.

\bibitem[{\citenamefont{Weng et~al.}(2015)\citenamefont{Weng, Fang, Fang,
  Bernevig, and Dai}}]{weng2015weyl}
\bibinfo{author}{\bibfnamefont{H.}~\bibnamefont{Weng}},
  \bibinfo{author}{\bibfnamefont{C.}~\bibnamefont{Fang}},
  \bibinfo{author}{\bibfnamefont{Z.}~\bibnamefont{Fang}},
  \bibinfo{author}{\bibfnamefont{B.~A.} \bibnamefont{Bernevig}},
  \bibnamefont{and} \bibinfo{author}{\bibfnamefont{X.}~\bibnamefont{Dai}},
  \bibinfo{journal}{Phys. Rev. X} \textbf{\bibinfo{volume}{5}},
  \bibinfo{pages}{011029} (\bibinfo{year}{2015}).

\bibitem[{\citenamefont{Qi and Zhang}(2011)}]{qi2011topological}
\bibinfo{author}{\bibfnamefont{X.-L.} \bibnamefont{Qi}} \bibnamefont{and}
  \bibinfo{author}{\bibfnamefont{S.-C.} \bibnamefont{Zhang}},
  \bibinfo{journal}{Rev. Mod. Phys.} \textbf{\bibinfo{volume}{83}},
  \bibinfo{pages}{1057} (\bibinfo{year}{2011}).

\bibitem[{\citenamefont{Soluyanov et~al.}(2015)\citenamefont{Soluyanov, Gresch,
  Wang, Wu, Troyer, Dai, and Bernevig}}]{soluyanov2015type}
\bibinfo{author}{\bibfnamefont{A.~A.} \bibnamefont{Soluyanov}},
  \bibinfo{author}{\bibfnamefont{D.}~\bibnamefont{Gresch}},
  \bibinfo{author}{\bibfnamefont{Z.}~\bibnamefont{Wang}},
  \bibinfo{author}{\bibfnamefont{Q.}~\bibnamefont{Wu}},
  \bibinfo{author}{\bibfnamefont{M.}~\bibnamefont{Troyer}},
  \bibinfo{author}{\bibfnamefont{X.}~\bibnamefont{Dai}}, \bibnamefont{and}
  \bibinfo{author}{\bibfnamefont{B.~A.} \bibnamefont{Bernevig}},
  \bibinfo{journal}{Nature (London)} \textbf{\bibinfo{volume}{527}},
  \bibinfo{pages}{495} (\bibinfo{year}{2015}), ISSN \bibinfo{issn}{1476-4687}.

\bibitem[{\citenamefont{Liu et~al.}(2018)\citenamefont{Liu, Sun, Kumar,
  Muechler, Sun, Jiao, Yang, Liu, Liang, Xu et~al.}}]{liu2018giant}
\bibinfo{author}{\bibfnamefont{E.}~\bibnamefont{Liu}},
  \bibinfo{author}{\bibfnamefont{Y.}~\bibnamefont{Sun}},
  \bibinfo{author}{\bibfnamefont{N.}~\bibnamefont{Kumar}},
  \bibinfo{author}{\bibfnamefont{L.}~\bibnamefont{Muechler}},
  \bibinfo{author}{\bibfnamefont{A.}~\bibnamefont{Sun}},
  \bibinfo{author}{\bibfnamefont{L.}~\bibnamefont{Jiao}},
  \bibinfo{author}{\bibfnamefont{S.-Y.} \bibnamefont{Yang}},
  \bibinfo{author}{\bibfnamefont{D.}~\bibnamefont{Liu}},
  \bibinfo{author}{\bibfnamefont{A.}~\bibnamefont{Liang}},
  \bibinfo{author}{\bibfnamefont{Q.}~\bibnamefont{Xu}}, \bibnamefont{et~al.},
  \bibinfo{journal}{Nat. Phys.} \textbf{\bibinfo{volume}{14}},
  \bibinfo{pages}{1125} (\bibinfo{year}{2018}), ISSN \bibinfo{issn}{1745-2481}.

\bibitem[{\citenamefont{Wang et~al.}(2018)\citenamefont{Wang, Xu, Lou, Liu, Li,
  Huang, Shen, Weng, Wang, and Lei}}]{wang2018large}
\bibinfo{author}{\bibfnamefont{Q.}~\bibnamefont{Wang}},
  \bibinfo{author}{\bibfnamefont{Y.}~\bibnamefont{Xu}},
  \bibinfo{author}{\bibfnamefont{R.}~\bibnamefont{Lou}},
  \bibinfo{author}{\bibfnamefont{Z.}~\bibnamefont{Liu}},
  \bibinfo{author}{\bibfnamefont{M.}~\bibnamefont{Li}},
  \bibinfo{author}{\bibfnamefont{Y.}~\bibnamefont{Huang}},
  \bibinfo{author}{\bibfnamefont{D.}~\bibnamefont{Shen}},
  \bibinfo{author}{\bibfnamefont{H.}~\bibnamefont{Weng}},
  \bibinfo{author}{\bibfnamefont{S.}~\bibnamefont{Wang}}, \bibnamefont{and}
  \bibinfo{author}{\bibfnamefont{H.}~\bibnamefont{Lei}}, \bibinfo{journal}{Nat.
  Commun.} \textbf{\bibinfo{volume}{9}}, \bibinfo{pages}{3681}
  (\bibinfo{year}{2018}), ISSN \bibinfo{issn}{2041-1723}.

\bibitem[{\citenamefont{Wang et~al.}(2021{\natexlab{a}})\citenamefont{Wang, Ge,
  Li, Liu, Xu, and Wang}}]{wang2021intrinsic}
\bibinfo{author}{\bibfnamefont{P.}~\bibnamefont{Wang}},
  \bibinfo{author}{\bibfnamefont{J.}~\bibnamefont{Ge}},
  \bibinfo{author}{\bibfnamefont{J.}~\bibnamefont{Li}},
  \bibinfo{author}{\bibfnamefont{Y.}~\bibnamefont{Liu}},
  \bibinfo{author}{\bibfnamefont{Y.}~\bibnamefont{Xu}}, \bibnamefont{and}
  \bibinfo{author}{\bibfnamefont{J.}~\bibnamefont{Wang}}, \bibinfo{journal}{The
  Innovation} \textbf{\bibinfo{volume}{2}}, \bibinfo{pages}{100098}
  (\bibinfo{year}{2021}{\natexlab{a}}), ISSN \bibinfo{issn}{2666-6758}.

\bibitem[{\citenamefont{Bernevig et~al.}(2022)\citenamefont{Bernevig, Felser,
  and Beidenkopf}}]{bernevig2022progress}
\bibinfo{author}{\bibfnamefont{B.~A.} \bibnamefont{Bernevig}},
  \bibinfo{author}{\bibfnamefont{C.}~\bibnamefont{Felser}}, \bibnamefont{and}
  \bibinfo{author}{\bibfnamefont{H.}~\bibnamefont{Beidenkopf}},
  \bibinfo{journal}{Nature (London)} \textbf{\bibinfo{volume}{603}},
  \bibinfo{pages}{41} (\bibinfo{year}{2022}), ISSN \bibinfo{issn}{1476-4687}.

\bibitem[{\citenamefont{Li et~al.}(2019)\citenamefont{Li, Li, Du, Wang, Gu,
  Zhang, He, Duan, and Xu}}]{li2019intrinsic}
\bibinfo{author}{\bibfnamefont{J.}~\bibnamefont{Li}},
  \bibinfo{author}{\bibfnamefont{Y.}~\bibnamefont{Li}},
  \bibinfo{author}{\bibfnamefont{S.}~\bibnamefont{Du}},
  \bibinfo{author}{\bibfnamefont{Z.}~\bibnamefont{Wang}},
  \bibinfo{author}{\bibfnamefont{B.-L.} \bibnamefont{Gu}},
  \bibinfo{author}{\bibfnamefont{S.-C.} \bibnamefont{Zhang}},
  \bibinfo{author}{\bibfnamefont{K.}~\bibnamefont{He}},
  \bibinfo{author}{\bibfnamefont{W.}~\bibnamefont{Duan}}, \bibnamefont{and}
  \bibinfo{author}{\bibfnamefont{Y.}~\bibnamefont{Xu}}, \bibinfo{journal}{Sci.
  Adv.} \textbf{\bibinfo{volume}{5}}, \bibinfo{pages}{eaaw5685}
  (\bibinfo{year}{2019}).

\bibitem[{\citenamefont{Liu et~al.}(2021{\natexlab{a}})\citenamefont{Liu, Yi,
  Wang, Shen, Xie, Yang, Fennel, Stuhr, Li, Weng et~al.}}]{liu2021anisotropic}
\bibinfo{author}{\bibfnamefont{C.}~\bibnamefont{Liu}},
  \bibinfo{author}{\bibfnamefont{C.}~\bibnamefont{Yi}},
  \bibinfo{author}{\bibfnamefont{X.}~\bibnamefont{Wang}},
  \bibinfo{author}{\bibfnamefont{J.}~\bibnamefont{Shen}},
  \bibinfo{author}{\bibfnamefont{T.}~\bibnamefont{Xie}},
  \bibinfo{author}{\bibfnamefont{L.}~\bibnamefont{Yang}},
  \bibinfo{author}{\bibfnamefont{T.}~\bibnamefont{Fennel}},
  \bibinfo{author}{\bibfnamefont{U.}~\bibnamefont{Stuhr}},
  \bibinfo{author}{\bibfnamefont{S.}~\bibnamefont{Li}},
  \bibinfo{author}{\bibfnamefont{H.}~\bibnamefont{Weng}}, \bibnamefont{et~al.},
  \bibinfo{journal}{Sci. China-Phys. Mech. Astron.}
  \textbf{\bibinfo{volume}{64}}, \bibinfo{pages}{257511}
  (\bibinfo{year}{2021}{\natexlab{a}}), ISSN \bibinfo{issn}{1869-1927}.

\bibitem[{\citenamefont{Guguchia et~al.}(2020)\citenamefont{Guguchia, Verezhak,
  Gawryluk, Tsirkin, Yin, Belopolski, Zhou, Simutis, Zhang, Cochran
  et~al.}}]{guguchia2020tunable}
\bibinfo{author}{\bibfnamefont{Z.}~\bibnamefont{Guguchia}},
  \bibinfo{author}{\bibfnamefont{J.~a.~T.} \bibnamefont{Verezhak}},
  \bibinfo{author}{\bibfnamefont{D.~J.} \bibnamefont{Gawryluk}},
  \bibinfo{author}{\bibfnamefont{S.~S.} \bibnamefont{Tsirkin}},
  \bibinfo{author}{\bibfnamefont{J.-X.} \bibnamefont{Yin}},
  \bibinfo{author}{\bibfnamefont{I.}~\bibnamefont{Belopolski}},
  \bibinfo{author}{\bibfnamefont{H.}~\bibnamefont{Zhou}},
  \bibinfo{author}{\bibfnamefont{G.}~\bibnamefont{Simutis}},
  \bibinfo{author}{\bibfnamefont{S.-S.} \bibnamefont{Zhang}},
  \bibinfo{author}{\bibfnamefont{T.~A.} \bibnamefont{Cochran}},
  \bibnamefont{et~al.}, \bibinfo{journal}{Nat. Commun.}
  \textbf{\bibinfo{volume}{11}}, \bibinfo{pages}{559} (\bibinfo{year}{2020}),
  ISSN \bibinfo{issn}{2041-1723}.

\bibitem[{\citenamefont{Yan}(2021)}]{yan2021weyl}
\bibinfo{author}{\bibfnamefont{B.}~\bibnamefont{Yan}}, \bibinfo{journal}{Sci.
  China-Phys. Mech. Astron.} \textbf{\bibinfo{volume}{64}},
  \bibinfo{pages}{217063} (\bibinfo{year}{2021}), ISSN
  \bibinfo{issn}{1869-1927}.

\bibitem[{\citenamefont{Liu et~al.}(2022)\citenamefont{Liu, Zhang, Han, and
  Liu}}]{liu2022chiral}
\bibinfo{author}{\bibfnamefont{P.}~\bibnamefont{Liu}},
  \bibinfo{author}{\bibfnamefont{A.}~\bibnamefont{Zhang}},
  \bibinfo{author}{\bibfnamefont{J.}~\bibnamefont{Han}}, \bibnamefont{and}
  \bibinfo{author}{\bibfnamefont{Q.}~\bibnamefont{Liu}}, \bibinfo{journal}{The
  Innovation} \textbf{\bibinfo{volume}{3}}, \bibinfo{pages}{100343}
  (\bibinfo{year}{2022}), ISSN \bibinfo{issn}{2666-6758}.

\bibitem[{\citenamefont{Otte et~al.}(2008)\citenamefont{Otte, Ternes, {von
  Bergmann}, Loth, Brune, Lutz, Hirjibehedin, and Heinrich}}]{otte2008role}
\bibinfo{author}{\bibfnamefont{A.~F.} \bibnamefont{Otte}},
  \bibinfo{author}{\bibfnamefont{M.}~\bibnamefont{Ternes}},
  \bibinfo{author}{\bibfnamefont{K.}~\bibnamefont{{von Bergmann}}},
  \bibinfo{author}{\bibfnamefont{S.}~\bibnamefont{Loth}},
  \bibinfo{author}{\bibfnamefont{H.}~\bibnamefont{Brune}},
  \bibinfo{author}{\bibfnamefont{C.~P.} \bibnamefont{Lutz}},
  \bibinfo{author}{\bibfnamefont{C.~F.} \bibnamefont{Hirjibehedin}},
  \bibnamefont{and} \bibinfo{author}{\bibfnamefont{A.~J.}
  \bibnamefont{Heinrich}}, \bibinfo{journal}{Nat. Phys.}
  \textbf{\bibinfo{volume}{4}}, \bibinfo{pages}{847} (\bibinfo{year}{2008}),
  ISSN \bibinfo{issn}{1745-2481}.

\bibitem[{\citenamefont{Akbari et~al.}(2009)\citenamefont{Akbari, Thalmeier,
  and Fulde}}]{akbari2009theory}
\bibinfo{author}{\bibfnamefont{A.}~\bibnamefont{Akbari}},
  \bibinfo{author}{\bibfnamefont{P.}~\bibnamefont{Thalmeier}},
  \bibnamefont{and} \bibinfo{author}{\bibfnamefont{P.}~\bibnamefont{Fulde}},
  \bibinfo{journal}{Phys. Rev. Lett.} \textbf{\bibinfo{volume}{102}},
  \bibinfo{pages}{106402} (\bibinfo{year}{2009}).

\bibitem[{\citenamefont{Fuhrman et~al.}(2015)\citenamefont{Fuhrman, Leiner,
  Nikoli{\'c}, Granroth, Stone, Lumsden, {DeBeer-Schmitt}, Alekseev, Mignot,
  Koohpayeh et~al.}}]{fuhrman2015interaction}
\bibinfo{author}{\bibfnamefont{W.~T.} \bibnamefont{Fuhrman}},
  \bibinfo{author}{\bibfnamefont{J.}~\bibnamefont{Leiner}},
  \bibinfo{author}{\bibfnamefont{P.}~\bibnamefont{Nikoli{\'c}}},
  \bibinfo{author}{\bibfnamefont{G.~E.} \bibnamefont{Granroth}},
  \bibinfo{author}{\bibfnamefont{M.~B.} \bibnamefont{Stone}},
  \bibinfo{author}{\bibfnamefont{M.~D.} \bibnamefont{Lumsden}},
  \bibinfo{author}{\bibfnamefont{L.}~\bibnamefont{{DeBeer-Schmitt}}},
  \bibinfo{author}{\bibfnamefont{P.~A.} \bibnamefont{Alekseev}},
  \bibinfo{author}{\bibfnamefont{J.-M.} \bibnamefont{Mignot}},
  \bibinfo{author}{\bibfnamefont{S.~M.} \bibnamefont{Koohpayeh}},
  \bibnamefont{et~al.}, \bibinfo{journal}{Phys. Rev. Lett.}
  \textbf{\bibinfo{volume}{114}}, \bibinfo{pages}{036401}
  (\bibinfo{year}{2015}).

\bibitem[{\citenamefont{Hillier et~al.}(2012)\citenamefont{Hillier, Adroja,
  Manuel, Anand, Taylor, McEwen, Rainford, and Koza}}]{hillier2012muon}
\bibinfo{author}{\bibfnamefont{A.~D.} \bibnamefont{Hillier}},
  \bibinfo{author}{\bibfnamefont{D.~T.} \bibnamefont{Adroja}},
  \bibinfo{author}{\bibfnamefont{P.}~\bibnamefont{Manuel}},
  \bibinfo{author}{\bibfnamefont{V.~K.} \bibnamefont{Anand}},
  \bibinfo{author}{\bibfnamefont{J.~W.} \bibnamefont{Taylor}},
  \bibinfo{author}{\bibfnamefont{K.~A.} \bibnamefont{McEwen}},
  \bibinfo{author}{\bibfnamefont{B.~D.} \bibnamefont{Rainford}},
  \bibnamefont{and} \bibinfo{author}{\bibfnamefont{M.~M.} \bibnamefont{Koza}},
  \bibinfo{journal}{Phys. Rev. B} \textbf{\bibinfo{volume}{85}},
  \bibinfo{pages}{134405} (\bibinfo{year}{2012}).

\bibitem[{\citenamefont{Iwasa et~al.}(2023)\citenamefont{Iwasa, Suyama,
  Ohira-Kawamura, Nakajima, Raymond, Steffens, Yamada, Matsuda, Aoki, Kawasaki
  et~al.}}]{iwasa2023weyl}
\bibinfo{author}{\bibfnamefont{K.}~\bibnamefont{Iwasa}},
  \bibinfo{author}{\bibfnamefont{K.}~\bibnamefont{Suyama}},
  \bibinfo{author}{\bibfnamefont{S.}~\bibnamefont{Ohira-Kawamura}},
  \bibinfo{author}{\bibfnamefont{K.}~\bibnamefont{Nakajima}},
  \bibinfo{author}{\bibfnamefont{S.}~\bibnamefont{Raymond}},
  \bibinfo{author}{\bibfnamefont{P.}~\bibnamefont{Steffens}},
  \bibinfo{author}{\bibfnamefont{A.}~\bibnamefont{Yamada}},
  \bibinfo{author}{\bibfnamefont{T.~D.} \bibnamefont{Matsuda}},
  \bibinfo{author}{\bibfnamefont{Y.}~\bibnamefont{Aoki}},
  \bibinfo{author}{\bibfnamefont{I.}~\bibnamefont{Kawasaki}},
  \bibnamefont{et~al.}, \bibinfo{journal}{Phy. Rev. Mater.}
  \textbf{\bibinfo{volume}{7}}, \bibinfo{pages}{014201} (\bibinfo{year}{2023}),
  \urlprefix\url{10.1103/PhysRevMaterials.7.014201}.

\bibitem[{\citenamefont{Thalmeier}(1984)}]{thalmeier1984theory}
\bibinfo{author}{\bibfnamefont{P.}~\bibnamefont{Thalmeier}},
  \bibinfo{journal}{J. Phys.: Condens. Matter} \textbf{\bibinfo{volume}{17}},
  \bibinfo{pages}{4153} (\bibinfo{year}{1984}),
  \urlprefix\url{https://doi.org/10.1088/0022-3719/17/23/015}.

\bibitem[{\citenamefont{Adroja et~al.}(2012)\citenamefont{Adroja, Del~Moral,
  De~La~Fuente, Fraile, Goremychkin, Taylor, Hillier, and
  Fernandez-Alonso}}]{adroja2012vibron}
\bibinfo{author}{\bibfnamefont{D.}~\bibnamefont{Adroja}},
  \bibinfo{author}{\bibfnamefont{A.}~\bibnamefont{Del~Moral}},
  \bibinfo{author}{\bibfnamefont{C.}~\bibnamefont{De~La~Fuente}},
  \bibinfo{author}{\bibfnamefont{A.}~\bibnamefont{Fraile}},
  \bibinfo{author}{\bibfnamefont{E.}~\bibnamefont{Goremychkin}},
  \bibinfo{author}{\bibfnamefont{J.}~\bibnamefont{Taylor}},
  \bibinfo{author}{\bibfnamefont{A.}~\bibnamefont{Hillier}}, \bibnamefont{and}
  \bibinfo{author}{\bibfnamefont{F.}~\bibnamefont{Fernandez-Alonso}},
  \bibinfo{journal}{Phy. Rev. Lett.} \textbf{\bibinfo{volume}{108}},
  \bibinfo{pages}{216402} (\bibinfo{year}{2012}),
  \urlprefix\url{https://doi.org/10.1103/PhysRevLett.108.216402}.

\bibitem[{\citenamefont{Bogdanov and
  Yablonskii}(1989)}]{bogdanov1989thermodynamically}
\bibinfo{author}{\bibfnamefont{A.~N.} \bibnamefont{Bogdanov}} \bibnamefont{and}
  \bibinfo{author}{\bibfnamefont{D.~A.} \bibnamefont{Yablonskii}},
  \bibinfo{journal}{Zh. Eksp. Teor. Fiz.} \textbf{\bibinfo{volume}{95}},
  \bibinfo{pages}{178} (\bibinfo{year}{1989}), \bibinfo{note}{[Sov. Phys. JETP
  \textbf{68}, 101 (1989)]}.

\bibitem[{\citenamefont{Kim et~al.}(2012)\citenamefont{Kim, Casa, Upton, Gog,
  Kim, Mitchell, {van Veenendaal}, Daghofer, {van den Brink}, Khaliullin
  et~al.}}]{kim2012magnetic}
\bibinfo{author}{\bibfnamefont{J.}~\bibnamefont{Kim}},
  \bibinfo{author}{\bibfnamefont{D.}~\bibnamefont{Casa}},
  \bibinfo{author}{\bibfnamefont{M.~H.} \bibnamefont{Upton}},
  \bibinfo{author}{\bibfnamefont{T.}~\bibnamefont{Gog}},
  \bibinfo{author}{\bibfnamefont{Y.-J.} \bibnamefont{Kim}},
  \bibinfo{author}{\bibfnamefont{J.~F.} \bibnamefont{Mitchell}},
  \bibinfo{author}{\bibfnamefont{M.}~\bibnamefont{{van Veenendaal}}},
  \bibinfo{author}{\bibfnamefont{M.}~\bibnamefont{Daghofer}},
  \bibinfo{author}{\bibfnamefont{J.}~\bibnamefont{{van den Brink}}},
  \bibinfo{author}{\bibfnamefont{G.}~\bibnamefont{Khaliullin}},
  \bibnamefont{et~al.}, \bibinfo{journal}{Phys. Rev. Lett.}
  \textbf{\bibinfo{volume}{108}}, \bibinfo{pages}{177003}
  (\bibinfo{year}{2012}).

\bibitem[{\citenamefont{Itoh et~al.}(2016)\citenamefont{Itoh, Endoh, Yokoo,
  Ibuka, Park, Kaneko, Takahashi, Tokura, and Nagaosa}}]{itoh2016weyl}
\bibinfo{author}{\bibfnamefont{S.}~\bibnamefont{Itoh}},
  \bibinfo{author}{\bibfnamefont{Y.}~\bibnamefont{Endoh}},
  \bibinfo{author}{\bibfnamefont{T.}~\bibnamefont{Yokoo}},
  \bibinfo{author}{\bibfnamefont{S.}~\bibnamefont{Ibuka}},
  \bibinfo{author}{\bibfnamefont{J.-G.} \bibnamefont{Park}},
  \bibinfo{author}{\bibfnamefont{Y.}~\bibnamefont{Kaneko}},
  \bibinfo{author}{\bibfnamefont{K.~S.} \bibnamefont{Takahashi}},
  \bibinfo{author}{\bibfnamefont{Y.}~\bibnamefont{Tokura}}, \bibnamefont{and}
  \bibinfo{author}{\bibfnamefont{N.}~\bibnamefont{Nagaosa}},
  \bibinfo{journal}{Nat. Commun.} \textbf{\bibinfo{volume}{7}},
  \bibinfo{pages}{11788} (\bibinfo{year}{2016}), ISSN
  \bibinfo{issn}{2041-1723}.

\bibitem[{\citenamefont{Jenni et~al.}(2019)\citenamefont{Jenni,
  Kunkem{\"o}ller, Br{\"u}ning, Lorenz, Sidis, Schneidewind, Nugroho, Rosch,
  Khomskii, and Braden}}]{jenni2019interplay}
\bibinfo{author}{\bibfnamefont{K.}~\bibnamefont{Jenni}},
  \bibinfo{author}{\bibfnamefont{S.}~\bibnamefont{Kunkem{\"o}ller}},
  \bibinfo{author}{\bibfnamefont{D.}~\bibnamefont{Br{\"u}ning}},
  \bibinfo{author}{\bibfnamefont{T.}~\bibnamefont{Lorenz}},
  \bibinfo{author}{\bibfnamefont{Y.}~\bibnamefont{Sidis}},
  \bibinfo{author}{\bibfnamefont{A.}~\bibnamefont{Schneidewind}},
  \bibinfo{author}{\bibfnamefont{A.~A.} \bibnamefont{Nugroho}},
  \bibinfo{author}{\bibfnamefont{A.}~\bibnamefont{Rosch}},
  \bibinfo{author}{\bibfnamefont{D.~I.} \bibnamefont{Khomskii}},
  \bibnamefont{and} \bibinfo{author}{\bibfnamefont{M.}~\bibnamefont{Braden}},
  \bibinfo{journal}{Phys. Rev. Lett.} \textbf{\bibinfo{volume}{123}},
  \bibinfo{pages}{017202} (\bibinfo{year}{2019}).

\bibitem[{\citenamefont{Cai et~al.}(2020)\citenamefont{Cai, Bao, Wang, Ma,
  Dong, Shangguan, Wang, Ran, Li, Kamazawa et~al.}}]{cai2020spin}
\bibinfo{author}{\bibfnamefont{Z.}~\bibnamefont{Cai}},
  \bibinfo{author}{\bibfnamefont{S.}~\bibnamefont{Bao}},
  \bibinfo{author}{\bibfnamefont{W.}~\bibnamefont{Wang}},
  \bibinfo{author}{\bibfnamefont{Z.}~\bibnamefont{Ma}},
  \bibinfo{author}{\bibfnamefont{Z.-Y.} \bibnamefont{Dong}},
  \bibinfo{author}{\bibfnamefont{Y.}~\bibnamefont{Shangguan}},
  \bibinfo{author}{\bibfnamefont{J.}~\bibnamefont{Wang}},
  \bibinfo{author}{\bibfnamefont{K.}~\bibnamefont{Ran}},
  \bibinfo{author}{\bibfnamefont{S.}~\bibnamefont{Li}},
  \bibinfo{author}{\bibfnamefont{K.}~\bibnamefont{Kamazawa}},
  \bibnamefont{et~al.}, \bibinfo{journal}{Phys. Rev. B}
  \textbf{\bibinfo{volume}{101}}, \bibinfo{pages}{134408}
  (\bibinfo{year}{2020}).

\bibitem[{\citenamefont{Liu et~al.}(2021{\natexlab{b}})\citenamefont{Liu, Shen,
  Gao, Yi, Liu, Xie, Yang, Danilkin, Deng, Wang et~al.}}]{liu2021spin}
\bibinfo{author}{\bibfnamefont{C.}~\bibnamefont{Liu}},
  \bibinfo{author}{\bibfnamefont{J.}~\bibnamefont{Shen}},
  \bibinfo{author}{\bibfnamefont{J.}~\bibnamefont{Gao}},
  \bibinfo{author}{\bibfnamefont{C.}~\bibnamefont{Yi}},
  \bibinfo{author}{\bibfnamefont{D.}~\bibnamefont{Liu}},
  \bibinfo{author}{\bibfnamefont{T.}~\bibnamefont{Xie}},
  \bibinfo{author}{\bibfnamefont{L.}~\bibnamefont{Yang}},
  \bibinfo{author}{\bibfnamefont{S.}~\bibnamefont{Danilkin}},
  \bibinfo{author}{\bibfnamefont{G.}~\bibnamefont{Deng}},
  \bibinfo{author}{\bibfnamefont{W.}~\bibnamefont{Wang}}, \bibnamefont{et~al.},
  \bibinfo{journal}{Sci. China-Phys. Mech. Astron.}
  \textbf{\bibinfo{volume}{64}}, \bibinfo{pages}{217062}
  (\bibinfo{year}{2021}{\natexlab{b}}), ISSN \bibinfo{issn}{1869-1927}.

\bibitem[{\citenamefont{Zhang et~al.}(2023)\citenamefont{Zhang, Deng, Sheng,
  Liu, Kumar, Shimada, Jiang, Liu, Shen, Li et~al.}}]{zhang2023chiral}
\bibinfo{author}{\bibfnamefont{A.}~\bibnamefont{Zhang}},
  \bibinfo{author}{\bibfnamefont{K.}~\bibnamefont{Deng}},
  \bibinfo{author}{\bibfnamefont{J.}~\bibnamefont{Sheng}},
  \bibinfo{author}{\bibfnamefont{P.}~\bibnamefont{Liu}},
  \bibinfo{author}{\bibfnamefont{S.}~\bibnamefont{Kumar}},
  \bibinfo{author}{\bibfnamefont{K.}~\bibnamefont{Shimada}},
  \bibinfo{author}{\bibfnamefont{Z.}~\bibnamefont{Jiang}},
  \bibinfo{author}{\bibfnamefont{Z.}~\bibnamefont{Liu}},
  \bibinfo{author}{\bibfnamefont{D.}~\bibnamefont{Shen}},
  \bibinfo{author}{\bibfnamefont{J.}~\bibnamefont{Li}}, \bibnamefont{et~al.},
  \bibinfo{journal}{Chin. Phys. Lett.} \textbf{\bibinfo{volume}{40}},
  \bibinfo{pages}{126101} (\bibinfo{year}{2023}), ISSN
  \bibinfo{issn}{0256-307X}.

\bibitem[{\citenamefont{Liu et~al.}(2023)\citenamefont{Liu, Lyu, Liu, Zhang,
  Yang, Du, Wang, Wei, and Liu}}]{liu2023structural}
\bibinfo{author}{\bibfnamefont{Y.}~\bibnamefont{Liu}},
  \bibinfo{author}{\bibfnamefont{M.}~\bibnamefont{Lyu}},
  \bibinfo{author}{\bibfnamefont{J.}~\bibnamefont{Liu}},
  \bibinfo{author}{\bibfnamefont{S.}~\bibnamefont{Zhang}},
  \bibinfo{author}{\bibfnamefont{J.}~\bibnamefont{Yang}},
  \bibinfo{author}{\bibfnamefont{Z.}~\bibnamefont{Du}},
  \bibinfo{author}{\bibfnamefont{B.}~\bibnamefont{Wang}},
  \bibinfo{author}{\bibfnamefont{H.}~\bibnamefont{Wei}}, \bibnamefont{and}
  \bibinfo{author}{\bibfnamefont{E.}~\bibnamefont{Liu}},
  \bibinfo{journal}{Chin. Phys. Lett.} \textbf{\bibinfo{volume}{40}},
  \bibinfo{pages}{047102} (\bibinfo{year}{2023}), ISSN
  \bibinfo{issn}{0256-307X}.

\bibitem[{\citenamefont{Zou et~al.}(2025)\citenamefont{Zou, Bai, Dai, Huang,
  and Niu}}]{zou2024experimentally}
\bibinfo{author}{\bibfnamefont{X.}~\bibnamefont{Zou}},
  \bibinfo{author}{\bibfnamefont{Y.}~\bibnamefont{Bai}},
  \bibinfo{author}{\bibfnamefont{Y.}~\bibnamefont{Dai}},
  \bibinfo{author}{\bibfnamefont{B.}~\bibnamefont{Huang}}, \bibnamefont{and}
  \bibinfo{author}{\bibfnamefont{C.}~\bibnamefont{Niu}},
  \bibinfo{journal}{TIMS} \textbf{\bibinfo{volume}{3}}, \bibinfo{pages}{100109}
  (\bibinfo{year}{2025}), ISSN \bibinfo{issn}{2959-8737}.

\bibitem[{\citenamefont{Yang et~al.}(2023)\citenamefont{Yang, Zhang, Sun,
  Felser, and Li}}]{yang2023topological}
\bibinfo{author}{\bibfnamefont{Q.}~\bibnamefont{Yang}},
  \bibinfo{author}{\bibfnamefont{Y.}~\bibnamefont{Zhang}},
  \bibinfo{author}{\bibfnamefont{Y.}~\bibnamefont{Sun}},
  \bibinfo{author}{\bibfnamefont{C.}~\bibnamefont{Felser}}, \bibnamefont{and}
  \bibinfo{author}{\bibfnamefont{G.}~\bibnamefont{Li}},
  \bibinfo{journal}{Innov. Mater.} \textbf{\bibinfo{volume}{1}},
  \bibinfo{pages}{100013} (\bibinfo{year}{2023}), ISSN
  \bibinfo{issn}{2959-8737}.

\bibitem[{\citenamefont{Su et~al.}(2021)\citenamefont{Su, Shi, Yuan, Wan,
  Cheng, Xi, Pi, Wang, Zou, Yu et~al.}}]{su2021multiple}
\bibinfo{author}{\bibfnamefont{H.}~\bibnamefont{Su}},
  \bibinfo{author}{\bibfnamefont{X.}~\bibnamefont{Shi}},
  \bibinfo{author}{\bibfnamefont{J.}~\bibnamefont{Yuan}},
  \bibinfo{author}{\bibfnamefont{Y.}~\bibnamefont{Wan}},
  \bibinfo{author}{\bibfnamefont{E.}~\bibnamefont{Cheng}},
  \bibinfo{author}{\bibfnamefont{C.}~\bibnamefont{Xi}},
  \bibinfo{author}{\bibfnamefont{L.}~\bibnamefont{Pi}},
  \bibinfo{author}{\bibfnamefont{X.}~\bibnamefont{Wang}},
  \bibinfo{author}{\bibfnamefont{Z.}~\bibnamefont{Zou}},
  \bibinfo{author}{\bibfnamefont{N.}~\bibnamefont{Yu}}, \bibnamefont{et~al.},
  \bibinfo{journal}{Phys. Rev. B} \textbf{\bibinfo{volume}{103}},
  \bibinfo{pages}{165128} (\bibinfo{year}{2021}).

\bibitem[{\citenamefont{Zhang et~al.}(2020)\citenamefont{Zhang, Wang, Pang,
  Han, Shang, Hung, Liu, Li, Saito, and Huang}}]{zhang2020anisotropic}
\bibinfo{author}{\bibfnamefont{K.}~\bibnamefont{Zhang}},
  \bibinfo{author}{\bibfnamefont{T.}~\bibnamefont{Wang}},
  \bibinfo{author}{\bibfnamefont{X.}~\bibnamefont{Pang}},
  \bibinfo{author}{\bibfnamefont{F.}~\bibnamefont{Han}},
  \bibinfo{author}{\bibfnamefont{S.-L.} \bibnamefont{Shang}},
  \bibinfo{author}{\bibfnamefont{N.~T.} \bibnamefont{Hung}},
  \bibinfo{author}{\bibfnamefont{Z.-K.} \bibnamefont{Liu}},
  \bibinfo{author}{\bibfnamefont{M.}~\bibnamefont{Li}},
  \bibinfo{author}{\bibfnamefont{R.}~\bibnamefont{Saito}}, \bibnamefont{and}
  \bibinfo{author}{\bibfnamefont{S.}~\bibnamefont{Huang}},
  \bibinfo{journal}{Phys. Rev. B} \textbf{\bibinfo{volume}{102}},
  \bibinfo{pages}{235162} (\bibinfo{year}{2020}).

\bibitem[{\citenamefont{Xu et~al.}(2017)\citenamefont{Xu, Alidoust, Chang, Lu,
  Singh, Belopolski, Sanchez, Zhang, Bian, Zheng et~al.}}]{xu2017discovery}
\bibinfo{author}{\bibfnamefont{S.-Y.} \bibnamefont{Xu}},
  \bibinfo{author}{\bibfnamefont{N.}~\bibnamefont{Alidoust}},
  \bibinfo{author}{\bibfnamefont{G.}~\bibnamefont{Chang}},
  \bibinfo{author}{\bibfnamefont{H.}~\bibnamefont{Lu}},
  \bibinfo{author}{\bibfnamefont{B.}~\bibnamefont{Singh}},
  \bibinfo{author}{\bibfnamefont{I.}~\bibnamefont{Belopolski}},
  \bibinfo{author}{\bibfnamefont{D.~S.} \bibnamefont{Sanchez}},
  \bibinfo{author}{\bibfnamefont{X.}~\bibnamefont{Zhang}},
  \bibinfo{author}{\bibfnamefont{G.}~\bibnamefont{Bian}},
  \bibinfo{author}{\bibfnamefont{H.}~\bibnamefont{Zheng}},
  \bibnamefont{et~al.}, \bibinfo{journal}{Sci. Adv.}
  \textbf{\bibinfo{volume}{3}}, \bibinfo{pages}{e1603266}
  (\bibinfo{year}{2017}).

\bibitem[{\citenamefont{Ng et~al.}(2021)\citenamefont{Ng, Luo, Yuan, Wu, Yang,
  and Shen}}]{ng2021origin}
\bibinfo{author}{\bibfnamefont{T.}~\bibnamefont{Ng}},
  \bibinfo{author}{\bibfnamefont{Y.}~\bibnamefont{Luo}},
  \bibinfo{author}{\bibfnamefont{J.}~\bibnamefont{Yuan}},
  \bibinfo{author}{\bibfnamefont{Y.}~\bibnamefont{Wu}},
  \bibinfo{author}{\bibfnamefont{H.}~\bibnamefont{Yang}}, \bibnamefont{and}
  \bibinfo{author}{\bibfnamefont{L.}~\bibnamefont{Shen}},
  \bibinfo{journal}{Phys. Rev. B} \textbf{\bibinfo{volume}{104}},
  \bibinfo{pages}{014412} (\bibinfo{year}{2021}).

\bibitem[{\citenamefont{Cao et~al.}(2022)\citenamefont{Cao, Zhao, Pei, Wang,
  Zhang, Ying, Zhao, Gao, Li, Yu et~al.}}]{cao2022pressure}
\bibinfo{author}{\bibfnamefont{W.}~\bibnamefont{Cao}},
  \bibinfo{author}{\bibfnamefont{N.}~\bibnamefont{Zhao}},
  \bibinfo{author}{\bibfnamefont{C.}~\bibnamefont{Pei}},
  \bibinfo{author}{\bibfnamefont{Q.}~\bibnamefont{Wang}},
  \bibinfo{author}{\bibfnamefont{Q.}~\bibnamefont{Zhang}},
  \bibinfo{author}{\bibfnamefont{T.}~\bibnamefont{Ying}},
  \bibinfo{author}{\bibfnamefont{Y.}~\bibnamefont{Zhao}},
  \bibinfo{author}{\bibfnamefont{L.}~\bibnamefont{Gao}},
  \bibinfo{author}{\bibfnamefont{C.}~\bibnamefont{Li}},
  \bibinfo{author}{\bibfnamefont{N.}~\bibnamefont{Yu}}, \bibnamefont{et~al.},
  \bibinfo{journal}{Phys. Rev. B} \textbf{\bibinfo{volume}{105}},
  \bibinfo{pages}{174502} (\bibinfo{year}{2022}).

\bibitem[{\citenamefont{Chang et~al.}(2018)\citenamefont{Chang, Singh, Xu,
  Bian, Huang, Hsu, Belopolski, Alidoust, Sanchez, Zheng
  et~al.}}]{chang2018magnetic}
\bibinfo{author}{\bibfnamefont{G.}~\bibnamefont{Chang}},
  \bibinfo{author}{\bibfnamefont{B.}~\bibnamefont{Singh}},
  \bibinfo{author}{\bibfnamefont{S.-Y.} \bibnamefont{Xu}},
  \bibinfo{author}{\bibfnamefont{G.}~\bibnamefont{Bian}},
  \bibinfo{author}{\bibfnamefont{S.-M.} \bibnamefont{Huang}},
  \bibinfo{author}{\bibfnamefont{C.-H.} \bibnamefont{Hsu}},
  \bibinfo{author}{\bibfnamefont{I.}~\bibnamefont{Belopolski}},
  \bibinfo{author}{\bibfnamefont{N.}~\bibnamefont{Alidoust}},
  \bibinfo{author}{\bibfnamefont{D.~S.} \bibnamefont{Sanchez}},
  \bibinfo{author}{\bibfnamefont{H.}~\bibnamefont{Zheng}},
  \bibnamefont{et~al.}, \bibinfo{journal}{Phys. Rev. B}
  \textbf{\bibinfo{volume}{97}}, \bibinfo{pages}{041104}
  (\bibinfo{year}{2018}).

\bibitem[{\citenamefont{Yang et~al.}(2021)\citenamefont{Yang, Singh, Gaudet,
  Lu, Huang, Chiu, Huang, Wang, Bahrami, Xu et~al.}}]{yang2021noncollinear}
\bibinfo{author}{\bibfnamefont{H.-Y.} \bibnamefont{Yang}},
  \bibinfo{author}{\bibfnamefont{B.}~\bibnamefont{Singh}},
  \bibinfo{author}{\bibfnamefont{J.}~\bibnamefont{Gaudet}},
  \bibinfo{author}{\bibfnamefont{B.}~\bibnamefont{Lu}},
  \bibinfo{author}{\bibfnamefont{C.-Y.} \bibnamefont{Huang}},
  \bibinfo{author}{\bibfnamefont{W.-C.} \bibnamefont{Chiu}},
  \bibinfo{author}{\bibfnamefont{S.-M.} \bibnamefont{Huang}},
  \bibinfo{author}{\bibfnamefont{B.}~\bibnamefont{Wang}},
  \bibinfo{author}{\bibfnamefont{F.}~\bibnamefont{Bahrami}},
  \bibinfo{author}{\bibfnamefont{B.}~\bibnamefont{Xu}}, \bibnamefont{et~al.},
  \bibinfo{journal}{Phys. Rev. B} \textbf{\bibinfo{volume}{103}},
  \bibinfo{pages}{115143} (\bibinfo{year}{2021}).

\bibitem[{\citenamefont{Bouaziz et~al.}(2024)\citenamefont{Bouaziz, Bihlmayer,
  Patrick, Staunton, and Bl{\"u}gel}}]{bouaziz2024origin}
\bibinfo{author}{\bibfnamefont{J.}~\bibnamefont{Bouaziz}},
  \bibinfo{author}{\bibfnamefont{G.}~\bibnamefont{Bihlmayer}},
  \bibinfo{author}{\bibfnamefont{C.~E.} \bibnamefont{Patrick}},
  \bibinfo{author}{\bibfnamefont{J.~B.} \bibnamefont{Staunton}},
  \bibnamefont{and}
  \bibinfo{author}{\bibfnamefont{S.}~\bibnamefont{Bl{\"u}gel}},
  \bibinfo{journal}{Phys. Rev. B} \textbf{\bibinfo{volume}{109}},
  \bibinfo{pages}{L201108} (\bibinfo{year}{2024}).

\bibitem[{\citenamefont{Piva et~al.}(2023)\citenamefont{Piva, Souza,
  {Brousseau-Couture}, Sorn, Pakuszewski, John, Adriano, C{\^o}t{\'e},
  Pagliuso, Paramekanti et~al.}}]{piva2023topological}
\bibinfo{author}{\bibfnamefont{M.~M.} \bibnamefont{Piva}},
  \bibinfo{author}{\bibfnamefont{J.~C.} \bibnamefont{Souza}},
  \bibinfo{author}{\bibfnamefont{V.}~\bibnamefont{{Brousseau-Couture}}},
  \bibinfo{author}{\bibfnamefont{S.}~\bibnamefont{Sorn}},
  \bibinfo{author}{\bibfnamefont{K.~R.} \bibnamefont{Pakuszewski}},
  \bibinfo{author}{\bibfnamefont{J.~K.} \bibnamefont{John}},
  \bibinfo{author}{\bibfnamefont{C.}~\bibnamefont{Adriano}},
  \bibinfo{author}{\bibfnamefont{M.}~\bibnamefont{C{\^o}t{\'e}}},
  \bibinfo{author}{\bibfnamefont{P.~G.} \bibnamefont{Pagliuso}},
  \bibinfo{author}{\bibfnamefont{A.}~\bibnamefont{Paramekanti}},
  \bibnamefont{et~al.}, \bibinfo{journal}{Phys. Rev. Res.}
  \textbf{\bibinfo{volume}{5}}, \bibinfo{pages}{013068} (\bibinfo{year}{2023}).

\bibitem[{\citenamefont{Sun et~al.}(2021)\citenamefont{Sun, Lee, Yang,
  Torchinsky, Tafti, and Orenstein}}]{sun2021mapping}
\bibinfo{author}{\bibfnamefont{Y.}~\bibnamefont{Sun}},
  \bibinfo{author}{\bibfnamefont{C.}~\bibnamefont{Lee}},
  \bibinfo{author}{\bibfnamefont{H.-Y.} \bibnamefont{Yang}},
  \bibinfo{author}{\bibfnamefont{D.~H.} \bibnamefont{Torchinsky}},
  \bibinfo{author}{\bibfnamefont{F.}~\bibnamefont{Tafti}}, \bibnamefont{and}
  \bibinfo{author}{\bibfnamefont{J.}~\bibnamefont{Orenstein}},
  \bibinfo{journal}{Phys. Rev. B} \textbf{\bibinfo{volume}{104}},
  \bibinfo{pages}{235119} (\bibinfo{year}{2021}).

\bibitem[{\citenamefont{Tzschaschel et~al.}(2024)\citenamefont{Tzschaschel,
  Qiu, Gao, Li, Guo, Yang, Zhang, Xie, Liu, Gao
  et~al.}}]{tzschaschel2024nonlinear}
\bibinfo{author}{\bibfnamefont{C.}~\bibnamefont{Tzschaschel}},
  \bibinfo{author}{\bibfnamefont{J.-X.} \bibnamefont{Qiu}},
  \bibinfo{author}{\bibfnamefont{X.-J.} \bibnamefont{Gao}},
  \bibinfo{author}{\bibfnamefont{H.-C.} \bibnamefont{Li}},
  \bibinfo{author}{\bibfnamefont{C.}~\bibnamefont{Guo}},
  \bibinfo{author}{\bibfnamefont{H.-Y.} \bibnamefont{Yang}},
  \bibinfo{author}{\bibfnamefont{C.-P.} \bibnamefont{Zhang}},
  \bibinfo{author}{\bibfnamefont{Y.-M.} \bibnamefont{Xie}},
  \bibinfo{author}{\bibfnamefont{Y.-F.} \bibnamefont{Liu}},
  \bibinfo{author}{\bibfnamefont{A.}~\bibnamefont{Gao}}, \bibnamefont{et~al.},
  \bibinfo{journal}{Nat. Commun.} \textbf{\bibinfo{volume}{15}},
  \bibinfo{pages}{3017} (\bibinfo{year}{2024}), ISSN \bibinfo{issn}{2041-1723}.

\bibitem[{\citenamefont{Cheng et~al.}(2024)\citenamefont{Cheng, Yan, Shi, Lou,
  Fedorov, Behnami, Yuan, Yang, Wang, Cheng et~al.}}]{cheng2024tunable}
\bibinfo{author}{\bibfnamefont{E.}~\bibnamefont{Cheng}},
  \bibinfo{author}{\bibfnamefont{L.}~\bibnamefont{Yan}},
  \bibinfo{author}{\bibfnamefont{X.}~\bibnamefont{Shi}},
  \bibinfo{author}{\bibfnamefont{R.}~\bibnamefont{Lou}},
  \bibinfo{author}{\bibfnamefont{A.}~\bibnamefont{Fedorov}},
  \bibinfo{author}{\bibfnamefont{M.}~\bibnamefont{Behnami}},
  \bibinfo{author}{\bibfnamefont{J.}~\bibnamefont{Yuan}},
  \bibinfo{author}{\bibfnamefont{P.}~\bibnamefont{Yang}},
  \bibinfo{author}{\bibfnamefont{B.}~\bibnamefont{Wang}},
  \bibinfo{author}{\bibfnamefont{J.-G.} \bibnamefont{Cheng}},
  \bibnamefont{et~al.}, \bibinfo{journal}{Nat. Commun.}
  \textbf{\bibinfo{volume}{15}}, \bibinfo{pages}{1467} (\bibinfo{year}{2024}),
  ISSN \bibinfo{issn}{2041-1723}.

\bibitem[{\citenamefont{Sakhya et~al.}(2023)\citenamefont{Sakhya, Huang,
  Dhakal, Gao, Regmi, Wang, Wen, He, Yao, Smith
  et~al.}}]{sakhya2023observation}
\bibinfo{author}{\bibfnamefont{A.~P.} \bibnamefont{Sakhya}},
  \bibinfo{author}{\bibfnamefont{C.-Y.} \bibnamefont{Huang}},
  \bibinfo{author}{\bibfnamefont{G.}~\bibnamefont{Dhakal}},
  \bibinfo{author}{\bibfnamefont{X.-J.} \bibnamefont{Gao}},
  \bibinfo{author}{\bibfnamefont{S.}~\bibnamefont{Regmi}},
  \bibinfo{author}{\bibfnamefont{B.}~\bibnamefont{Wang}},
  \bibinfo{author}{\bibfnamefont{W.}~\bibnamefont{Wen}},
  \bibinfo{author}{\bibfnamefont{R.-H.} \bibnamefont{He}},
  \bibinfo{author}{\bibfnamefont{X.}~\bibnamefont{Yao}},
  \bibinfo{author}{\bibfnamefont{R.}~\bibnamefont{Smith}},
  \bibnamefont{et~al.}, \bibinfo{journal}{Phys. Rev. Mater.}
  \textbf{\bibinfo{volume}{7}}, \bibinfo{pages}{L051202}
  (\bibinfo{year}{2023}).

\bibitem[{\citenamefont{Morita et~al.}(2025)\citenamefont{Morita, Nakanishi,
  Iwata, Ohwada, Nishioka, Kousa, Nurmamat, Yamagami, Kimura, Yamada
  et~al.}}]{morita2024zone}
\bibinfo{author}{\bibfnamefont{Y.}~\bibnamefont{Morita}},
  \bibinfo{author}{\bibfnamefont{K.}~\bibnamefont{Nakanishi}},
  \bibinfo{author}{\bibfnamefont{T.}~\bibnamefont{Iwata}},
  \bibinfo{author}{\bibfnamefont{K.}~\bibnamefont{Ohwada}},
  \bibinfo{author}{\bibfnamefont{Y.}~\bibnamefont{Nishioka}},
  \bibinfo{author}{\bibfnamefont{T.}~\bibnamefont{Kousa}},
  \bibinfo{author}{\bibfnamefont{M.}~\bibnamefont{Nurmamat}},
  \bibinfo{author}{\bibfnamefont{K.}~\bibnamefont{Yamagami}},
  \bibinfo{author}{\bibfnamefont{A.}~\bibnamefont{Kimura}},
  \bibinfo{author}{\bibfnamefont{T.}~\bibnamefont{Yamada}},
  \bibnamefont{et~al.}, \bibinfo{journal}{Phys. Rev. B}
  \textbf{\bibinfo{volume}{111}}, \bibinfo{pages}{L081116}
  (\bibinfo{year}{2025}).

\bibitem[{\citenamefont{Lou et~al.}(2023)\citenamefont{Lou, Fedorov, Zhao,
  Yaresko, B{\"u}chner, and Borisenko}}]{lou2023signature}
\bibinfo{author}{\bibfnamefont{R.}~\bibnamefont{Lou}},
  \bibinfo{author}{\bibfnamefont{A.}~\bibnamefont{Fedorov}},
  \bibinfo{author}{\bibfnamefont{L.}~\bibnamefont{Zhao}},
  \bibinfo{author}{\bibfnamefont{A.}~\bibnamefont{Yaresko}},
  \bibinfo{author}{\bibfnamefont{B.}~\bibnamefont{B{\"u}chner}},
  \bibnamefont{and}
  \bibinfo{author}{\bibfnamefont{S.}~\bibnamefont{Borisenko}},
  \bibinfo{journal}{Phys. Rev. B} \textbf{\bibinfo{volume}{107}},
  \bibinfo{pages}{035158} (\bibinfo{year}{2023}).

\bibitem[{\citenamefont{Lyu et~al.}(2020{\natexlab{a}})\citenamefont{Lyu, Wang,
  Ramesh~Kumar, Zhao, Xiang, and Sun}}]{lyu2020large}
\bibinfo{author}{\bibfnamefont{M.}~\bibnamefont{Lyu}},
  \bibinfo{author}{\bibfnamefont{Z.}~\bibnamefont{Wang}},
  \bibinfo{author}{\bibfnamefont{K.}~\bibnamefont{Ramesh~Kumar}},
  \bibinfo{author}{\bibfnamefont{H.}~\bibnamefont{Zhao}},
  \bibinfo{author}{\bibfnamefont{J.}~\bibnamefont{Xiang}}, \bibnamefont{and}
  \bibinfo{author}{\bibfnamefont{P.}~\bibnamefont{Sun}}, \bibinfo{journal}{J.
  Appl. Phys.} \textbf{\bibinfo{volume}{127}}, \bibinfo{pages}{193903}
  (\bibinfo{year}{2020}{\natexlab{a}}), ISSN \bibinfo{issn}{0021-8979}.

\bibitem[{\citenamefont{Wu et~al.}(2023)\citenamefont{Wu, Chi, Zuo, Xu, Zhao,
  Luo, and Zhu}}]{wu2023field}
\bibinfo{author}{\bibfnamefont{L.}~\bibnamefont{Wu}},
  \bibinfo{author}{\bibfnamefont{S.}~\bibnamefont{Chi}},
  \bibinfo{author}{\bibfnamefont{H.}~\bibnamefont{Zuo}},
  \bibinfo{author}{\bibfnamefont{G.}~\bibnamefont{Xu}},
  \bibinfo{author}{\bibfnamefont{L.}~\bibnamefont{Zhao}},
  \bibinfo{author}{\bibfnamefont{Y.}~\bibnamefont{Luo}}, \bibnamefont{and}
  \bibinfo{author}{\bibfnamefont{Z.}~\bibnamefont{Zhu}}, \bibinfo{journal}{npj
  Quantum Mater.} \textbf{\bibinfo{volume}{8}}, \bibinfo{pages}{4}
  (\bibinfo{year}{2023}), ISSN \bibinfo{issn}{2397-4648}.

\bibitem[{\citenamefont{Lyu et~al.}(2020{\natexlab{b}})\citenamefont{Lyu,
  Xiang, Mi, Zhao, Wang, Liu, Chen, Ren, Li, and Sun}}]{lyu2020nonsaturating}
\bibinfo{author}{\bibfnamefont{M.}~\bibnamefont{Lyu}},
  \bibinfo{author}{\bibfnamefont{J.}~\bibnamefont{Xiang}},
  \bibinfo{author}{\bibfnamefont{Z.}~\bibnamefont{Mi}},
  \bibinfo{author}{\bibfnamefont{H.}~\bibnamefont{Zhao}},
  \bibinfo{author}{\bibfnamefont{Z.}~\bibnamefont{Wang}},
  \bibinfo{author}{\bibfnamefont{E.}~\bibnamefont{Liu}},
  \bibinfo{author}{\bibfnamefont{G.}~\bibnamefont{Chen}},
  \bibinfo{author}{\bibfnamefont{Z.}~\bibnamefont{Ren}},
  \bibinfo{author}{\bibfnamefont{G.}~\bibnamefont{Li}}, \bibnamefont{and}
  \bibinfo{author}{\bibfnamefont{P.}~\bibnamefont{Sun}},
  \bibinfo{journal}{Phys. Rev. B} \textbf{\bibinfo{volume}{102}},
  \bibinfo{pages}{085143} (\bibinfo{year}{2020}{\natexlab{b}}).

\bibitem[{\citenamefont{Yang et~al.}(2020)\citenamefont{Yang, Singh, Lu, Huang,
  Bahrami, Chiu, Graf, Huang, Wang, Lin et~al.}}]{yang2020transition}
\bibinfo{author}{\bibfnamefont{H.-Y.} \bibnamefont{Yang}},
  \bibinfo{author}{\bibfnamefont{B.}~\bibnamefont{Singh}},
  \bibinfo{author}{\bibfnamefont{B.}~\bibnamefont{Lu}},
  \bibinfo{author}{\bibfnamefont{C.-Y.} \bibnamefont{Huang}},
  \bibinfo{author}{\bibfnamefont{F.}~\bibnamefont{Bahrami}},
  \bibinfo{author}{\bibfnamefont{W.-C.} \bibnamefont{Chiu}},
  \bibinfo{author}{\bibfnamefont{D.}~\bibnamefont{Graf}},
  \bibinfo{author}{\bibfnamefont{S.-M.} \bibnamefont{Huang}},
  \bibinfo{author}{\bibfnamefont{B.}~\bibnamefont{Wang}},
  \bibinfo{author}{\bibfnamefont{H.}~\bibnamefont{Lin}}, \bibnamefont{et~al.},
  \bibinfo{journal}{APL Mater.} \textbf{\bibinfo{volume}{8}},
  \bibinfo{pages}{011111} (\bibinfo{year}{2020}), ISSN
  \bibinfo{issn}{2166-532X}.

\bibitem[{\citenamefont{Wang et~al.}(2023)\citenamefont{Wang, Dong, Huang,
  Wang, Guo, Wang, Ren, Li, Sun, Dai et~al.}}]{wang2023quantum}
\bibinfo{author}{\bibfnamefont{J.-F.} \bibnamefont{Wang}},
  \bibinfo{author}{\bibfnamefont{Q.-X.} \bibnamefont{Dong}},
  \bibinfo{author}{\bibfnamefont{Y.-F.} \bibnamefont{Huang}},
  \bibinfo{author}{\bibfnamefont{Z.-S.} \bibnamefont{Wang}},
  \bibinfo{author}{\bibfnamefont{Z.-P.} \bibnamefont{Guo}},
  \bibinfo{author}{\bibfnamefont{Z.-J.} \bibnamefont{Wang}},
  \bibinfo{author}{\bibfnamefont{Z.-A.} \bibnamefont{Ren}},
  \bibinfo{author}{\bibfnamefont{G.}~\bibnamefont{Li}},
  \bibinfo{author}{\bibfnamefont{P.-J.} \bibnamefont{Sun}},
  \bibinfo{author}{\bibfnamefont{X.}~\bibnamefont{Dai}}, \bibnamefont{et~al.},
  \bibinfo{journal}{Phys. Rev. B} \textbf{\bibinfo{volume}{108}},
  \bibinfo{pages}{024423} (\bibinfo{year}{2023}).

\bibitem[{\citenamefont{Yamada et~al.}(2024)\citenamefont{Yamada, Nomoto,
  Miyake, Terakawa, Kikkawa, Arita, Tokunaga, Taguchi, Tokura, and
  Hirschberger}}]{yamada2024nernst}
\bibinfo{author}{\bibfnamefont{R.}~\bibnamefont{Yamada}},
  \bibinfo{author}{\bibfnamefont{T.}~\bibnamefont{Nomoto}},
  \bibinfo{author}{\bibfnamefont{A.}~\bibnamefont{Miyake}},
  \bibinfo{author}{\bibfnamefont{T.}~\bibnamefont{Terakawa}},
  \bibinfo{author}{\bibfnamefont{A.}~\bibnamefont{Kikkawa}},
  \bibinfo{author}{\bibfnamefont{R.}~\bibnamefont{Arita}},
  \bibinfo{author}{\bibfnamefont{M.}~\bibnamefont{Tokunaga}},
  \bibinfo{author}{\bibfnamefont{Y.}~\bibnamefont{Taguchi}},
  \bibinfo{author}{\bibfnamefont{Y.}~\bibnamefont{Tokura}}, \bibnamefont{and}
  \bibinfo{author}{\bibfnamefont{M.}~\bibnamefont{Hirschberger}},
  \bibinfo{journal}{Phys. Rev. X} \textbf{\bibinfo{volume}{14}},
  \bibinfo{pages}{021012} (\bibinfo{year}{2024}).

\bibitem[{\citenamefont{Kunze et~al.}(2024)\citenamefont{Kunze, K{\"o}pf, Cao,
  Qi, and Kuntscher}}]{kunze2024optical}
\bibinfo{author}{\bibfnamefont{J.}~\bibnamefont{Kunze}},
  \bibinfo{author}{\bibfnamefont{M.}~\bibnamefont{K{\"o}pf}},
  \bibinfo{author}{\bibfnamefont{W.}~\bibnamefont{Cao}},
  \bibinfo{author}{\bibfnamefont{Y.}~\bibnamefont{Qi}}, \bibnamefont{and}
  \bibinfo{author}{\bibfnamefont{C.~A.} \bibnamefont{Kuntscher}},
  \bibinfo{journal}{Phys. Rev. B} \textbf{\bibinfo{volume}{109}},
  \bibinfo{pages}{195130} (\bibinfo{year}{2024}).

\bibitem[{\citenamefont{Wang et~al.}(2021{\natexlab{b}})\citenamefont{Wang,
  Guo, and Wang}}]{wang2021structure}
\bibinfo{author}{\bibfnamefont{T.}~\bibnamefont{Wang}},
  \bibinfo{author}{\bibfnamefont{Y.}~\bibnamefont{Guo}}, \bibnamefont{and}
  \bibinfo{author}{\bibfnamefont{C.}~\bibnamefont{Wang}},
  \bibinfo{journal}{Chin. Phys. B} \textbf{\bibinfo{volume}{30}},
  \bibinfo{pages}{075102} (\bibinfo{year}{2021}{\natexlab{b}}), ISSN
  \bibinfo{issn}{1674-1056}.

\bibitem[{\citenamefont{Wang et~al.}(2024)\citenamefont{Wang, Zhen, Meng,
  Plokhikh, Wu, Gawryluk, Xu, Zhan, Shi, Pomjakushina et~al.}}]{wang2024spin}
\bibinfo{author}{\bibfnamefont{Y.}~\bibnamefont{Wang}},
  \bibinfo{author}{\bibfnamefont{Z.}~\bibnamefont{Zhen}},
  \bibinfo{author}{\bibfnamefont{J.}~\bibnamefont{Meng}},
  \bibinfo{author}{\bibfnamefont{I.}~\bibnamefont{Plokhikh}},
  \bibinfo{author}{\bibfnamefont{D.}~\bibnamefont{Wu}},
  \bibinfo{author}{\bibfnamefont{D.~J.} \bibnamefont{Gawryluk}},
  \bibinfo{author}{\bibfnamefont{Y.}~\bibnamefont{Xu}},
  \bibinfo{author}{\bibfnamefont{Q.}~\bibnamefont{Zhan}},
  \bibinfo{author}{\bibfnamefont{M.}~\bibnamefont{Shi}},
  \bibinfo{author}{\bibfnamefont{E.}~\bibnamefont{Pomjakushina}},
  \bibnamefont{et~al.}, \bibinfo{journal}{Sci. China-Phys. Mech. Astron.}
  \textbf{\bibinfo{volume}{67}}, \bibinfo{pages}{107512}
  (\bibinfo{year}{2024}), ISSN \bibinfo{issn}{1869-1927}.

\bibitem[{\citenamefont{Suzuki et~al.}(2019)\citenamefont{Suzuki, Savary, Liu,
  Lynn, Balents, and Checkelsky}}]{suzuki2019singular}
\bibinfo{author}{\bibfnamefont{T.}~\bibnamefont{Suzuki}},
  \bibinfo{author}{\bibfnamefont{L.}~\bibnamefont{Savary}},
  \bibinfo{author}{\bibfnamefont{J.-P.} \bibnamefont{Liu}},
  \bibinfo{author}{\bibfnamefont{J.~W.} \bibnamefont{Lynn}},
  \bibinfo{author}{\bibfnamefont{L.}~\bibnamefont{Balents}}, \bibnamefont{and}
  \bibinfo{author}{\bibfnamefont{J.~G.} \bibnamefont{Checkelsky}},
  \bibinfo{journal}{Science} \textbf{\bibinfo{volume}{365}},
  \bibinfo{pages}{377} (\bibinfo{year}{2019}).

\bibitem[{\citenamefont{Puphal et~al.}(2020)\citenamefont{Puphal, Pomjakushin,
  Kanazawa, Ukleev, Gawryluk, Ma, Naamneh, Plumb, Keller, Cubitt
  et~al.}}]{puphal2020topological}
\bibinfo{author}{\bibfnamefont{P.}~\bibnamefont{Puphal}},
  \bibinfo{author}{\bibfnamefont{V.}~\bibnamefont{Pomjakushin}},
  \bibinfo{author}{\bibfnamefont{N.}~\bibnamefont{Kanazawa}},
  \bibinfo{author}{\bibfnamefont{V.}~\bibnamefont{Ukleev}},
  \bibinfo{author}{\bibfnamefont{D.~J.} \bibnamefont{Gawryluk}},
  \bibinfo{author}{\bibfnamefont{J.}~\bibnamefont{Ma}},
  \bibinfo{author}{\bibfnamefont{M.}~\bibnamefont{Naamneh}},
  \bibinfo{author}{\bibfnamefont{N.~C.} \bibnamefont{Plumb}},
  \bibinfo{author}{\bibfnamefont{L.}~\bibnamefont{Keller}},
  \bibinfo{author}{\bibfnamefont{R.}~\bibnamefont{Cubitt}},
  \bibnamefont{et~al.}, \bibinfo{journal}{Phys. Rev. Lett.}
  \textbf{\bibinfo{volume}{124}}, \bibinfo{pages}{017202}
  (\bibinfo{year}{2020}).

\bibitem[{\citenamefont{Drucker et~al.}(2023)\citenamefont{Drucker, Nguyen,
  Han, Siriviboon, Luo, Andrejevic, Zhu, Bednik, Nguyen, Chen
  et~al.}}]{drucker2023topology}
\bibinfo{author}{\bibfnamefont{N.~C.} \bibnamefont{Drucker}},
  \bibinfo{author}{\bibfnamefont{T.}~\bibnamefont{Nguyen}},
  \bibinfo{author}{\bibfnamefont{F.}~\bibnamefont{Han}},
  \bibinfo{author}{\bibfnamefont{P.}~\bibnamefont{Siriviboon}},
  \bibinfo{author}{\bibfnamefont{X.}~\bibnamefont{Luo}},
  \bibinfo{author}{\bibfnamefont{N.}~\bibnamefont{Andrejevic}},
  \bibinfo{author}{\bibfnamefont{Z.}~\bibnamefont{Zhu}},
  \bibinfo{author}{\bibfnamefont{G.}~\bibnamefont{Bednik}},
  \bibinfo{author}{\bibfnamefont{Q.~T.} \bibnamefont{Nguyen}},
  \bibinfo{author}{\bibfnamefont{Z.}~\bibnamefont{Chen}}, \bibnamefont{et~al.},
  \bibinfo{journal}{Nat. Commun.} \textbf{\bibinfo{volume}{14}},
  \bibinfo{pages}{5182} (\bibinfo{year}{2023}), ISSN \bibinfo{issn}{2041-1723}.

\bibitem[{\citenamefont{He et~al.}(2023)\citenamefont{He, Li, Zeng, Zhu, Tan,
  Zhang, Cao, and Luo}}]{he2023pressure}
\bibinfo{author}{\bibfnamefont{X.}~\bibnamefont{He}},
  \bibinfo{author}{\bibfnamefont{Y.}~\bibnamefont{Li}},
  \bibinfo{author}{\bibfnamefont{H.}~\bibnamefont{Zeng}},
  \bibinfo{author}{\bibfnamefont{Z.}~\bibnamefont{Zhu}},
  \bibinfo{author}{\bibfnamefont{S.}~\bibnamefont{Tan}},
  \bibinfo{author}{\bibfnamefont{Y.}~\bibnamefont{Zhang}},
  \bibinfo{author}{\bibfnamefont{C.}~\bibnamefont{Cao}}, \bibnamefont{and}
  \bibinfo{author}{\bibfnamefont{Y.}~\bibnamefont{Luo}}, \bibinfo{journal}{Sci.
  China-Phys. Mech. Astron.} \textbf{\bibinfo{volume}{66}},
  \bibinfo{pages}{237011} (\bibinfo{year}{2023}), ISSN
  \bibinfo{issn}{1869-1927}.

\bibitem[{\citenamefont{Sanchez et~al.}(2020)\citenamefont{Sanchez, Chang,
  Belopolski, Lu, Yin, Alidoust, Xu, Cochran, Zhang, Bian
  et~al.}}]{sanchez2020observation}
\bibinfo{author}{\bibfnamefont{D.~S.} \bibnamefont{Sanchez}},
  \bibinfo{author}{\bibfnamefont{G.}~\bibnamefont{Chang}},
  \bibinfo{author}{\bibfnamefont{I.}~\bibnamefont{Belopolski}},
  \bibinfo{author}{\bibfnamefont{H.}~\bibnamefont{Lu}},
  \bibinfo{author}{\bibfnamefont{J.-X.} \bibnamefont{Yin}},
  \bibinfo{author}{\bibfnamefont{N.}~\bibnamefont{Alidoust}},
  \bibinfo{author}{\bibfnamefont{X.}~\bibnamefont{Xu}},
  \bibinfo{author}{\bibfnamefont{T.~A.} \bibnamefont{Cochran}},
  \bibinfo{author}{\bibfnamefont{X.}~\bibnamefont{Zhang}},
  \bibinfo{author}{\bibfnamefont{Y.}~\bibnamefont{Bian}}, \bibnamefont{et~al.},
  \bibinfo{journal}{Nat. Commun.} \textbf{\bibinfo{volume}{11}},
  \bibinfo{pages}{3356} (\bibinfo{year}{2020}), ISSN \bibinfo{issn}{2041-1723}.

\bibitem[{\citenamefont{Meng et~al.}(2019)\citenamefont{Meng, Wu, Qiu, Wang,
  Liu, Xia, Yuan, Chang, and Tian}}]{meng2019large}
\bibinfo{author}{\bibfnamefont{B.}~\bibnamefont{Meng}},
  \bibinfo{author}{\bibfnamefont{H.}~\bibnamefont{Wu}},
  \bibinfo{author}{\bibfnamefont{Y.}~\bibnamefont{Qiu}},
  \bibinfo{author}{\bibfnamefont{C.}~\bibnamefont{Wang}},
  \bibinfo{author}{\bibfnamefont{Y.}~\bibnamefont{Liu}},
  \bibinfo{author}{\bibfnamefont{Z.}~\bibnamefont{Xia}},
  \bibinfo{author}{\bibfnamefont{S.}~\bibnamefont{Yuan}},
  \bibinfo{author}{\bibfnamefont{H.}~\bibnamefont{Chang}}, \bibnamefont{and}
  \bibinfo{author}{\bibfnamefont{Z.}~\bibnamefont{Tian}}, \bibinfo{journal}{APL
  Mater.} \textbf{\bibinfo{volume}{7}}, \bibinfo{pages}{051110}
  (\bibinfo{year}{2019}), ISSN \bibinfo{issn}{2166-532X}.

\bibitem[{\citenamefont{Gaudet et~al.}(2021)\citenamefont{Gaudet, Yang, Baidya,
  Lu, Xu, Zhao, {Rodriguez-Rivera}, Hoffmann, Graf, Torchinsky
  et~al.}}]{gaudet2021weyl}
\bibinfo{author}{\bibfnamefont{J.}~\bibnamefont{Gaudet}},
  \bibinfo{author}{\bibfnamefont{H.-Y.} \bibnamefont{Yang}},
  \bibinfo{author}{\bibfnamefont{S.}~\bibnamefont{Baidya}},
  \bibinfo{author}{\bibfnamefont{B.}~\bibnamefont{Lu}},
  \bibinfo{author}{\bibfnamefont{G.}~\bibnamefont{Xu}},
  \bibinfo{author}{\bibfnamefont{Y.}~\bibnamefont{Zhao}},
  \bibinfo{author}{\bibfnamefont{J.~A.} \bibnamefont{{Rodriguez-Rivera}}},
  \bibinfo{author}{\bibfnamefont{C.~M.} \bibnamefont{Hoffmann}},
  \bibinfo{author}{\bibfnamefont{D.~E.} \bibnamefont{Graf}},
  \bibinfo{author}{\bibfnamefont{D.~H.} \bibnamefont{Torchinsky}},
  \bibnamefont{et~al.}, \bibinfo{journal}{Nat. Mater.}
  \textbf{\bibinfo{volume}{20}}, \bibinfo{pages}{1650} (\bibinfo{year}{2021}),
  ISSN \bibinfo{issn}{1476-4660}.

\bibitem[{\citenamefont{Wang et~al.}(2022)\citenamefont{Wang, Dong, Guo, Lv,
  Huang, Xiang, Ren, Wang, Sun, Li et~al.}}]{wang2022ndalsi}
\bibinfo{author}{\bibfnamefont{J.-F.} \bibnamefont{Wang}},
  \bibinfo{author}{\bibfnamefont{Q.-X.} \bibnamefont{Dong}},
  \bibinfo{author}{\bibfnamefont{Z.-P.} \bibnamefont{Guo}},
  \bibinfo{author}{\bibfnamefont{M.}~\bibnamefont{Lv}},
  \bibinfo{author}{\bibfnamefont{Y.-F.} \bibnamefont{Huang}},
  \bibinfo{author}{\bibfnamefont{J.-S.} \bibnamefont{Xiang}},
  \bibinfo{author}{\bibfnamefont{Z.-A.} \bibnamefont{Ren}},
  \bibinfo{author}{\bibfnamefont{Z.-J.} \bibnamefont{Wang}},
  \bibinfo{author}{\bibfnamefont{P.-J.} \bibnamefont{Sun}},
  \bibinfo{author}{\bibfnamefont{G.}~\bibnamefont{Li}}, \bibnamefont{et~al.},
  \bibinfo{journal}{Phys. Rev. B} \textbf{\bibinfo{volume}{105}},
  \bibinfo{pages}{144435} (\bibinfo{year}{2022}).

\bibitem[{\citenamefont{Lygouras et~al.}(2024)\citenamefont{Lygouras, Yang,
  Yao, Gaudet, Hao, Cao, {Rodriguez-Rivera}, Podlesnyak, Bl{\"u}gel,
  Nikoli{\'c} et~al.}}]{lygouras2024magnetic}
\bibinfo{author}{\bibfnamefont{C.~J.} \bibnamefont{Lygouras}},
  \bibinfo{author}{\bibfnamefont{H.-Y.} \bibnamefont{Yang}},
  \bibinfo{author}{\bibfnamefont{X.}~\bibnamefont{Yao}},
  \bibinfo{author}{\bibfnamefont{J.}~\bibnamefont{Gaudet}},
  \bibinfo{author}{\bibfnamefont{Y.}~\bibnamefont{Hao}},
  \bibinfo{author}{\bibfnamefont{H.}~\bibnamefont{Cao}},
  \bibinfo{author}{\bibfnamefont{J.~A.} \bibnamefont{{Rodriguez-Rivera}}},
  \bibinfo{author}{\bibfnamefont{A.}~\bibnamefont{Podlesnyak}},
  \bibinfo{author}{\bibfnamefont{S.}~\bibnamefont{Bl{\"u}gel}},
  \bibinfo{author}{\bibfnamefont{P.}~\bibnamefont{Nikoli{\'c}}},
  \bibnamefont{et~al.}, \emph{\bibinfo{title}{Magnetic excitations and
  interactions in the {{Weyl}} ferrimagnet {{NdAlSi}}}} (\bibinfo{year}{2024}),
  \eprint{2412.20743}.

\bibitem[{\citenamefont{Li et~al.}(2023)\citenamefont{Li, Zhang, Wang, Liu,
  Guo, Rienks, Chen, Bertran, Yang, Phuyal et~al.}}]{li2023emergence}
\bibinfo{author}{\bibfnamefont{C.}~\bibnamefont{Li}},
  \bibinfo{author}{\bibfnamefont{J.}~\bibnamefont{Zhang}},
  \bibinfo{author}{\bibfnamefont{Y.}~\bibnamefont{Wang}},
  \bibinfo{author}{\bibfnamefont{H.}~\bibnamefont{Liu}},
  \bibinfo{author}{\bibfnamefont{Q.}~\bibnamefont{Guo}},
  \bibinfo{author}{\bibfnamefont{E.}~\bibnamefont{Rienks}},
  \bibinfo{author}{\bibfnamefont{W.}~\bibnamefont{Chen}},
  \bibinfo{author}{\bibfnamefont{F.}~\bibnamefont{Bertran}},
  \bibinfo{author}{\bibfnamefont{H.}~\bibnamefont{Yang}},
  \bibinfo{author}{\bibfnamefont{D.}~\bibnamefont{Phuyal}},
  \bibnamefont{et~al.}, \bibinfo{journal}{Nat. Commun.}
  \textbf{\bibinfo{volume}{14}}, \bibinfo{pages}{7185} (\bibinfo{year}{2023}),
  ISSN \bibinfo{issn}{2041-1723}.

\bibitem[{\citenamefont{Zhang et~al.}(2024)\citenamefont{Zhang, Tu, Li, Tang,
  Nie, Li, Li, Qi, Wu, Zhou et~al.}}]{zhang2024abnormally}
\bibinfo{author}{\bibfnamefont{N.}~\bibnamefont{Zhang}},
  \bibinfo{author}{\bibfnamefont{D.}~\bibnamefont{Tu}},
  \bibinfo{author}{\bibfnamefont{D.}~\bibnamefont{Li}},
  \bibinfo{author}{\bibfnamefont{K.}~\bibnamefont{Tang}},
  \bibinfo{author}{\bibfnamefont{L.}~\bibnamefont{Nie}},
  \bibinfo{author}{\bibfnamefont{H.}~\bibnamefont{Li}},
  \bibinfo{author}{\bibfnamefont{H.}~\bibnamefont{Li}},
  \bibinfo{author}{\bibfnamefont{T.}~\bibnamefont{Qi}},
  \bibinfo{author}{\bibfnamefont{T.}~\bibnamefont{Wu}},
  \bibinfo{author}{\bibfnamefont{J.}~\bibnamefont{Zhou}}, \bibnamefont{et~al.},
  \bibinfo{journal}{Nat. Commun.} \textbf{\bibinfo{volume}{15}},
  \bibinfo{pages}{10255} (\bibinfo{year}{2024}), ISSN
  \bibinfo{issn}{2041-1723}.

\bibitem[{\citenamefont{Yao et~al.}(2023)\citenamefont{Yao, Gaudet, Verma,
  Graf, Yang, Bahrami, Zhang, Aczel, Subedi, Torchinsky et~al.}}]{yao2023large}
\bibinfo{author}{\bibfnamefont{X.}~\bibnamefont{Yao}},
  \bibinfo{author}{\bibfnamefont{J.}~\bibnamefont{Gaudet}},
  \bibinfo{author}{\bibfnamefont{R.}~\bibnamefont{Verma}},
  \bibinfo{author}{\bibfnamefont{D.~E.} \bibnamefont{Graf}},
  \bibinfo{author}{\bibfnamefont{H.-Y.} \bibnamefont{Yang}},
  \bibinfo{author}{\bibfnamefont{F.}~\bibnamefont{Bahrami}},
  \bibinfo{author}{\bibfnamefont{R.}~\bibnamefont{Zhang}},
  \bibinfo{author}{\bibfnamefont{A.~A.} \bibnamefont{Aczel}},
  \bibinfo{author}{\bibfnamefont{S.}~\bibnamefont{Subedi}},
  \bibinfo{author}{\bibfnamefont{D.~H.} \bibnamefont{Torchinsky}},
  \bibnamefont{et~al.}, \bibinfo{journal}{Phys. Rev. X}
  \textbf{\bibinfo{volume}{13}}, \bibinfo{pages}{011035}
  (\bibinfo{year}{2023}).

\bibitem[{\citenamefont{Xu et~al.}(2022)\citenamefont{Xu, Niu, Bai, Zhu, Yuan,
  He, Han, Zhao, Yang, Xia et~al.}}]{xu2022shubnikov}
\bibinfo{author}{\bibfnamefont{L.}~\bibnamefont{Xu}},
  \bibinfo{author}{\bibfnamefont{H.}~\bibnamefont{Niu}},
  \bibinfo{author}{\bibfnamefont{Y.}~\bibnamefont{Bai}},
  \bibinfo{author}{\bibfnamefont{H.}~\bibnamefont{Zhu}},
  \bibinfo{author}{\bibfnamefont{S.}~\bibnamefont{Yuan}},
  \bibinfo{author}{\bibfnamefont{X.}~\bibnamefont{He}},
  \bibinfo{author}{\bibfnamefont{Y.}~\bibnamefont{Han}},
  \bibinfo{author}{\bibfnamefont{L.}~\bibnamefont{Zhao}},
  \bibinfo{author}{\bibfnamefont{Y.}~\bibnamefont{Yang}},
  \bibinfo{author}{\bibfnamefont{Z.}~\bibnamefont{Xia}}, \bibnamefont{et~al.},
  \bibinfo{journal}{J. Phys. Condens. Matter} \textbf{\bibinfo{volume}{34}},
  \bibinfo{pages}{485701} (\bibinfo{year}{2022}),
  \urlprefix\url{https://doi.org/10.1088/1361-648X/ac987a}.

\bibitem[{\citenamefont{Walicka et~al.}(2024)\citenamefont{Walicka, Blacque,
  Gornicka, White, Klimczuk, and von Rohr}}]{walicka2024magnetism}
\bibinfo{author}{\bibfnamefont{D.~I.} \bibnamefont{Walicka}},
  \bibinfo{author}{\bibfnamefont{O.}~\bibnamefont{Blacque}},
  \bibinfo{author}{\bibfnamefont{K.}~\bibnamefont{Gornicka}},
  \bibinfo{author}{\bibfnamefont{J.~S.} \bibnamefont{White}},
  \bibinfo{author}{\bibfnamefont{T.}~\bibnamefont{Klimczuk}}, \bibnamefont{and}
  \bibinfo{author}{\bibfnamefont{F.~O.} \bibnamefont{von Rohr}},
  \bibinfo{journal}{arXiv:2412.12795}  (\bibinfo{year}{2024}),
  \urlprefix\url{https://doi.org/10.48550/arXiv.2412.12795}.

\bibitem[{\citenamefont{Nag et~al.}(2024)\citenamefont{Nag, Das, Bhowal,
  Nishioka, Bandyopadhyay, Sarker, Kumar, Kuroda, Gopalan, Kimura
  et~al.}}]{nag2024gdalsi}
\bibinfo{author}{\bibfnamefont{J.}~\bibnamefont{Nag}},
  \bibinfo{author}{\bibfnamefont{B.}~\bibnamefont{Das}},
  \bibinfo{author}{\bibfnamefont{S.}~\bibnamefont{Bhowal}},
  \bibinfo{author}{\bibfnamefont{Y.}~\bibnamefont{Nishioka}},
  \bibinfo{author}{\bibfnamefont{B.}~\bibnamefont{Bandyopadhyay}},
  \bibinfo{author}{\bibfnamefont{S.}~\bibnamefont{Sarker}},
  \bibinfo{author}{\bibfnamefont{S.}~\bibnamefont{Kumar}},
  \bibinfo{author}{\bibfnamefont{K.}~\bibnamefont{Kuroda}},
  \bibinfo{author}{\bibfnamefont{V.}~\bibnamefont{Gopalan}},
  \bibinfo{author}{\bibfnamefont{A.}~\bibnamefont{Kimura}},
  \bibnamefont{et~al.}, \bibinfo{journal}{Phys. Rev. B}
  \textbf{\bibinfo{volume}{110}}, \bibinfo{pages}{224436}
  (\bibinfo{year}{2024}),
  \urlprefix\url{https://link.aps.org/doi/10.1103/PhysRevB.110.224436}.

\bibitem[{\citenamefont{Gong et~al.}(2024)\citenamefont{Gong, Wang, Ma, Zeng,
  Lin, Han, Wang, and Xia}}]{gong2024magnetic}
\bibinfo{author}{\bibfnamefont{J.}~\bibnamefont{Gong}},
  \bibinfo{author}{\bibfnamefont{H.}~\bibnamefont{Wang}},
  \bibinfo{author}{\bibfnamefont{X.}~\bibnamefont{Ma}},
  \bibinfo{author}{\bibfnamefont{X.}~\bibnamefont{Zeng}},
  \bibinfo{author}{\bibfnamefont{J.}~\bibnamefont{Lin}},
  \bibinfo{author}{\bibfnamefont{K.}~\bibnamefont{Han}},
  \bibinfo{author}{\bibfnamefont{Y.}~\bibnamefont{Wang}}, \bibnamefont{and}
  \bibinfo{author}{\bibfnamefont{T.}~\bibnamefont{Xia}},
  \bibinfo{journal}{Chin. Phy. B} \textbf{\bibinfo{volume}{33}},
  \bibinfo{pages}{077302} (\bibinfo{year}{2024}),
  \urlprefix\url{https://doi.org/10.1088/1674-1056/ad41ba}.

\bibitem[{\citenamefont{Le et~al.}(2023)\citenamefont{Le, Guidi, Bewley,
  Stewart, Schooneveld, Raspino, Pooley, Boxall, Gascoyne, Rhodes
  et~al.}}]{LE2023168646}
\bibinfo{author}{\bibfnamefont{M.~D.} \bibnamefont{Le}},
  \bibinfo{author}{\bibfnamefont{T.}~\bibnamefont{Guidi}},
  \bibinfo{author}{\bibfnamefont{R.}~\bibnamefont{Bewley}},
  \bibinfo{author}{\bibfnamefont{J.~R.} \bibnamefont{Stewart}},
  \bibinfo{author}{\bibfnamefont{E.~M.} \bibnamefont{Schooneveld}},
  \bibinfo{author}{\bibfnamefont{D.}~\bibnamefont{Raspino}},
  \bibinfo{author}{\bibfnamefont{D.~E.} \bibnamefont{Pooley}},
  \bibinfo{author}{\bibfnamefont{J.}~\bibnamefont{Boxall}},
  \bibinfo{author}{\bibfnamefont{K.~F.} \bibnamefont{Gascoyne}},
  \bibinfo{author}{\bibfnamefont{N.~J.} \bibnamefont{Rhodes}},
  \bibnamefont{et~al.}, \bibinfo{journal}{Nucl. Instrum. Methods Phys. Res.
  Sect. A} \textbf{\bibinfo{volume}{1056}}, \bibinfo{pages}{168646}
  (\bibinfo{year}{2023}), ISSN \bibinfo{issn}{0168-9002}.

\bibitem[{\citenamefont{Ding et~al.}(2023)\citenamefont{Ding, Adroja, Aouane,
  Sun, Yang, and Luo}}]{ding2023crystal}
\bibinfo{author}{\bibfnamefont{J.}~\bibnamefont{Ding}},
  \bibinfo{author}{\bibfnamefont{D.}~\bibnamefont{Adroja}},
  \bibinfo{author}{\bibfnamefont{M.}~\bibnamefont{Aouane}},
  \bibinfo{author}{\bibfnamefont{Y.}~\bibnamefont{Sun}},
  \bibinfo{author}{\bibfnamefont{L.}~\bibnamefont{Yang}}, \bibnamefont{and}
  \bibinfo{author}{\bibfnamefont{H.-Q.} \bibnamefont{Luo}},
  \bibinfo{journal}{STFC ISIS Neutron and Muon Source}  (\bibinfo{year}{2023}),
  \urlprefix\url{https://doi.org/10.5286/ISIS.E.RB2310280}.

\bibitem[{\citenamefont{Fulde and Loewenhaupt}(1985)}]{fulde1985magnetic}
\bibinfo{author}{\bibfnamefont{P.}~\bibnamefont{Fulde}} \bibnamefont{and}
  \bibinfo{author}{\bibfnamefont{M.}~\bibnamefont{Loewenhaupt}},
  \bibinfo{journal}{Adv. Phys.} \textbf{\bibinfo{volume}{34}},
  \bibinfo{pages}{589} (\bibinfo{year}{1985}), ISSN \bibinfo{issn}{0001-8732}.

\bibitem[{\citenamefont{Jensen and Mackintosh}(1991)}]{Jensen1991}
\bibinfo{author}{\bibfnamefont{J.}~\bibnamefont{Jensen}} \bibnamefont{and}
  \bibinfo{author}{\bibfnamefont{A.~R.} \bibnamefont{Mackintosh}},
  \emph{\bibinfo{title}{Rare earth magnetism: structures and excitations}}
  (\bibinfo{publisher}{Oxford University, New York}, \bibinfo{year}{1991}),
  \urlprefix\url{https://doi.org/10.1093/oso/9780198520276.001.0001}.

\bibitem[{\citenamefont{Pang et~al.}(2022)\citenamefont{Pang, Peng, Fang, and
  Huang}}]{Pang2022}
\bibinfo{author}{\bibfnamefont{X.}~\bibnamefont{Pang}},
  \bibinfo{author}{\bibfnamefont{B.}~\bibnamefont{Peng}},
  \bibinfo{author}{\bibfnamefont{Y.}~\bibnamefont{Fang}}, \bibnamefont{and}
  \bibinfo{author}{\bibfnamefont{F.}~\bibnamefont{Huang}}, \bibinfo{journal}{J.
  Solid State Chem.} \textbf{\bibinfo{volume}{308}}, \bibinfo{pages}{122877}
  (\bibinfo{year}{2022}), ISSN \bibinfo{issn}{0022-4596}.

\bibitem[{\citenamefont{Stevens}(1952)}]{stevens1952matrix}
\bibinfo{author}{\bibfnamefont{K.~W.~H.} \bibnamefont{Stevens}},
  \bibinfo{journal}{Proc. Phys. Soc. A} \textbf{\bibinfo{volume}{65}},
  \bibinfo{pages}{209} (\bibinfo{year}{1952}), ISSN \bibinfo{issn}{0370-1298}.

\bibitem[{\citenamefont{Hutchings}(1964)}]{hutchings1964point}
\bibinfo{author}{\bibfnamefont{M.}~\bibnamefont{Hutchings}},
  \bibinfo{journal}{Solid State Phys.} \textbf{\bibinfo{volume}{16}},
  \bibinfo{pages}{227} (\bibinfo{year}{1964}).

\bibitem[{\citenamefont{Boothroyd}(2020)}]{boothroyd2020principles}
\bibinfo{author}{\bibfnamefont{A.~T.} \bibnamefont{Boothroyd}},
  \emph{\bibinfo{title}{Principles of Neutron Scattering from Condensed
  Matter}} (\bibinfo{publisher}{Oxford University, New York},
  \bibinfo{year}{2020}).

\bibitem[{\citenamefont{Arnold et~al.}(2014)\citenamefont{Arnold, Bilheux,
  Borreguero, Buts, Campbell, Chapon, Doucet, Draper, Ferraz~Leal, Gigg
  et~al.}}]{arnold2014mantid}
\bibinfo{author}{\bibfnamefont{O.}~\bibnamefont{Arnold}},
  \bibinfo{author}{\bibfnamefont{J.}~\bibnamefont{Bilheux}},
  \bibinfo{author}{\bibfnamefont{J.}~\bibnamefont{Borreguero}},
  \bibinfo{author}{\bibfnamefont{A.}~\bibnamefont{Buts}},
  \bibinfo{author}{\bibfnamefont{S.}~\bibnamefont{Campbell}},
  \bibinfo{author}{\bibfnamefont{L.}~\bibnamefont{Chapon}},
  \bibinfo{author}{\bibfnamefont{M.}~\bibnamefont{Doucet}},
  \bibinfo{author}{\bibfnamefont{N.}~\bibnamefont{Draper}},
  \bibinfo{author}{\bibfnamefont{R.}~\bibnamefont{Ferraz~Leal}},
  \bibinfo{author}{\bibfnamefont{M.}~\bibnamefont{Gigg}}, \bibnamefont{et~al.},
  \bibinfo{journal}{Nucl. Instrum. Methods Phys. Res., Sect. A}
  \textbf{\bibinfo{volume}{764}}, \bibinfo{pages}{156} (\bibinfo{year}{2014}),
  ISSN \bibinfo{issn}{0168-9002}.

\bibitem[{\citenamefont{Scheie}(2021)}]{scheie2021pycrystalfield}
\bibinfo{author}{\bibfnamefont{A.}~\bibnamefont{Scheie}}, \bibinfo{journal}{J.
  Appl. Crystallogr.} \textbf{\bibinfo{volume}{54}}, \bibinfo{pages}{356}
  (\bibinfo{year}{2021}), ISSN \bibinfo{issn}{1600-5767}.

\bibitem[{\citenamefont{Zhou et~al.}(2025)\citenamefont{Zhou, Chang, Yang, and
  Liang}}]{zhou2025weyl}
\bibinfo{author}{\bibfnamefont{M.}~\bibnamefont{Zhou}},
  \bibinfo{author}{\bibfnamefont{H.-R.} \bibnamefont{Chang}},
  \bibinfo{author}{\bibfnamefont{L.}~\bibnamefont{Yang}}, \bibnamefont{and}
  \bibinfo{author}{\bibfnamefont{L.}~\bibnamefont{Liang}},
  \emph{\bibinfo{title}{Weyl-mediated {{Ruderman-Kittel-Kasuya-Yosida}}
  interaction revisited: imaginary-time formalism and finite temperature
  effects}} (\bibinfo{year}{2025}), \eprint{2504.11052},
  \urlprefix\url{https://doi.org/10.48550/arXiv.2504.11052}.

\bibitem[{\citenamefont{Rotter}(2004)}]{rotter2004using}
\bibinfo{author}{\bibfnamefont{M.}~\bibnamefont{Rotter}}, \bibinfo{journal}{J.
  Magn. Magn. Mater.} \textbf{\bibinfo{volume}{272--276}},
  \bibinfo{pages}{E481} (\bibinfo{year}{2004}), ISSN \bibinfo{issn}{0304-8853}.

\bibitem[{\citenamefont{Luo et~al.}(2025)\citenamefont{Luo, Chen, Zhan, and
  Yu}}]{luo2025correlated}
\bibinfo{author}{\bibfnamefont{X.}~\bibnamefont{Luo}},
  \bibinfo{author}{\bibfnamefont{Y.-G.} \bibnamefont{Chen}},
  \bibinfo{author}{\bibfnamefont{Y.-M.} \bibnamefont{Zhan}}, \bibnamefont{and}
  \bibinfo{author}{\bibfnamefont{Y.}~\bibnamefont{Yu}},
  \emph{\bibinfo{title}{Correlated helimagnetic configuration in a
  nonsymmorphic magnetic nodal semimetal}} (\bibinfo{year}{2025}),
  \eprint{2502.03832},
  \urlprefix\url{https://doi.org/10.48550/arXiv.2502.03832}.

\end{thebibliography}

\end{document}